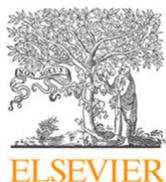

Contents lists available at ScienceDirect

# Physics Reports

journal homepage: www.elsevier.com/locate/physrep

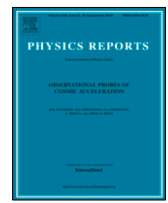

# Majorana quasiparticles in atomic spin chains on superconductors

Stephan Rachel [a], Roland Wiesendanger [b],*

[a] *School of Physics, University of Melbourne, Parkville, VIC 3010, Australia*
[b] *Department of Physics, University of Hamburg, Germany*

A B S T R A C T

For the past decade, Majorana quasiparticles have become one of the hot topics in condensed matter research. Besides the fundamental interest in the realization of particles being their own antiparticles, going back to basic concepts of elementary particle physics, Majorana quasiparticles in condensed matter systems offer exciting potential applications in topological quantum computation due to their non-Abelian quantum exchange statistics. Motivated by theoretical predictions about possible realizations of Majorana quasiparticles as zero-energy modes at boundaries of topological superconductors, experimental efforts have focussed in particular on quasi-one-dimensional semiconductor–superconductor and magnet–superconductor hybrid systems. However, an unambiguous proof of the existence of Majorana quasiparticles is still challenging and requires considerable improvements in materials science, atomic-scale characterization and control of interface quality, as well as complementary approaches of detecting various facets of Majorana quasiparticles. Bottom-up atom-by-atom fabrication of disorder-free atomic spin chains on atomically clean superconducting substrates has recently allowed deep insight into the emergence of topological sub-gap Shiba bands and associated Majorana states from the level of individual atoms up to extended chains, thereby offering the possibility for critical tests of Majorana physics in disorder-free model-type 1D hybrid systems.

## Contents

* Corresponding author.
   *E-mail address:* wiesendanger@physnet.uni-hamburg.de (R. Wiesendanger).

## 1. Introduction

Since the theoretical prediction of the existence of fermionic particles being their own antiparticles by Ettore Majorana [1] in 1937, there have been great efforts to detect such exotic particles in various subfields of physics [2]. While there is no consensus yet that a fermion exists being its own antiparticle in elementary particle physics, it has been shown theoretically that Majorana quasiparticles can be realized as zero-energy modes at boundaries of topological superconductors [3–23], which may be fabricated artificially by proximitizing semiconducting nanowires or atomic spin chains to $s$-wave superconductors. Besides the fundamental interest in proving the existence of a novel, exotic quasiparticle in a solid-state system, fault-tolerant quantum computing is the major motivation for hunting Majorana quasiparticles. In contrast to Majorana fermions in three-dimensional particle physics [2], these Majorana quasiparticles in effectively two-dimensional condensed matter systems exhibit a non-Abelian quantum exchange statistics [24–26] making them promising candidates to serve as robust qubits in topological quantum computation schemes [27–31]. Indeed, quantum-mechanical decoherence is the most serious limitation to rapid advances of quantum computers, therefore topologically protected, i.e., fault-tolerant Majorana qubits could potentially lead to a breakthrough of quantum computer technology.

Signatures of Majorana quasiparticles have first been reported about a decade ago based on heterostructures consisting of nano-scale semiconducting nanowires with high spin–orbit coupling, such as InAs or InSb, and elemental $s$-wave superconductors, such as Al, in the presence of a variable in-plane applied magnetic field [32–34]. Numerous follow-up publications have appeared based on the same or similar semiconductor–superconductor heterostructures probing additional signatures of Majorana quasiparticles [35–37], as theoretically predicted, such as the quantized value of conductance of the zero-energy boundary modes or exponentially decaying energy oscillations of the Majorana modes as a function of magnetic field strength. However, none of the experimental observations in this early period of Majorana quasiparticle research could rigorously prove the existence of these exotic quasiparticles [38–41]. Besides the need for distinguishing Majorana quasiparticles from trivial Andreev bound states at semiconductor–superconductor interfaces, disorder-related low-energy states associated with non-perfect interfaces can additionally complicate the interpretation of quantum transport measurements [42,43] typically used for the search of Majorana quasiparticles in semiconductor–superconductor heterostructures [38].

As an alternative platform for probing Majorana physics, atomic spin chains on superconductor substrates have been proposed theoretically [7–23]. Compared to semiconductor–superconductor heterostructures, no external magnetic field is required in this case. While an external magnetic field has the advantage as a tuning parameter, it might lead to experimental complications or challenges for future devices. Moreover, the electronic states of the spin chains interacting with a superconducting substrate can directly be accessed by scanning tunneling microscopy (STM) and spectroscopy (STS) techniques with high spatial and energy resolution [11]. Even the spin signature of Majorana states can be probed by spin-polarized STM (SP-STM) making use of spin-polarized tunneling into a highly spin-polarized STM probe tip [44,45]. Early attempts towards this direction were based on the self-assembly of ferromagnetic chains, made of Fe or Co, on superconducting substrates with high spin–orbit coupling, such as Pb(110) [46–51]. Indeed, signatures of Majorana bound states at the ends of those nanowires were found either as single features [46] or as 'double-eye' features [49]. In these early experiments, an unavoidable side-effect of the self-assembly process of nanowires was disorder due to intermixing between Fe and Pb as a result of the annealing treatment. The nanowires were also interrupted by adsorbed Fe clusters so that the simultaneous STM probing of both ends of the nanowires was prohibited.

More recently, the fabrication of well-defined and disorder-free spin chains on atomically clean superconductor substrates has become possible by STM-based single-atom manipulation techniques [52–59], which is the main focus of this review. Various kinds of magnetic transition-metal (Cr, Mn, Fe, Co) chains on different elemental superconducting substrates, including Re(0001), Nb(110), Ta(110), etc., have been constructed with variable interatomic distances along the spin chains and with different crystallographic orientations with respect to the superconducting substrate lattice. Moreover, the spin textures of these atomic magnetic chains could directly be revealed by atomic-resolution SP-STM measurements in dependence of the structural properties of the atomic-scale magnet–superconductor hybrid systems [52,55,57]. The new atomically precise model-type platform combined with innovative atomic-scale characterization tools nowadays allows for critical tests of various theoretically predicted signatures of Majorana quasiparticles. This is not only limited to the observation of zero-energy modes at the chain's ends [52,55,59], but includes the correlated appearance or disappearance of Majorana states at both ends of the atomic spin chains [52,57], oscillations of hybridizing Majorana modes as a function of chain length [57], and the direct observation of the emergence of topological sub-gap (Shiba) bands [55]. Most importantly, it is possible to directly probe the bulk-boundary correspondence [55], expected for a topological phase. Several of the experimental systems discussed in this Review clearly show topological properties and their zero-energy states are compatible with Majorana modes, although they might still suffer from a too small topological gap [55] or hybridization effects [57]. With the recent advances based on atomic-scale fabrication and probing technologies, which includes not only linear 1D chains [52–59], but also much more sophisticated model-type platforms such as tri-junctions [30,60], 2D atomic wire networks [61], and atomic rings [62,63], researchers are now well prepared for addressing the next challenge, i.e., controlled Majorana state manipulation.

## 2. Topological superconductivity and Majorana states in atomic spin chains on *s*-wave superconductors

The current experimental efforts are best understood by first discussing some of the theoretical foundations of topological superconductivity and Majorana quasiparticles. The Kitaev chain [3] serves as minimal model to describe a 1D topological superconductor and associated Majorana states at the chain's ends. For a given superconducting pairing

> **Box 1: Kitaev chain**
>
> (I) The Kitaev chain [3], a 1D topological superconductor model, features the most relevant low-energy properties of this class of models. It contains real nearest-neighbor hoppings, chemical potential and a nearest-neighbor pairing. For spinless fermions, such a pairing is of topological p-wave type.
>
> $$H = \sum_j \left( \underbrace{-t\, c_j^\dagger c_{j+1}}_{\text{Hopping}} - \underbrace{\frac{\mu}{2}\left(c_j^\dagger c_j - \tfrac{1}{2}\right)}_{\text{Chem. Potential}} + \underbrace{\Delta_p c_j c_{j+1}}_{\text{Pairing}} + \text{H.c.} \right)$$
>
> (II) Exact mapping from (complex-valued) Dirac fermion operators $c_j$ to (real-valued) Majorana fermion operators $\gamma_{2j}$:
>
> $$\gamma_{2j-1} = c_j + c_j^\dagger, \quad \gamma_{2j} = i\left(c_j^\dagger - c_j\right) \quad \text{and, correspondingly,} \quad c_j = \tfrac{1}{2}(\gamma_{2j-1} + i\gamma_{2j}), \quad c_j^\dagger = \tfrac{1}{2}(\gamma_{2j-1} - i\gamma_{2j})$$
>
> (III) Consider special limits to understand the different phases: (i) Trivial limit $|\Delta| = t = 0, \mu < 0$
>
> $$H = \frac{i}{2}(-\mu) \sum_j \gamma_{2j-1}\gamma_{2j}$$
>
> 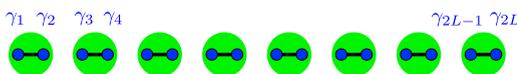
>
> (ii) Topological limit $|\Delta| = t > 0, \mu = 0$
>
> $$H = it \sum_j \gamma_{2j}\gamma_{2j+1}$$
>
> 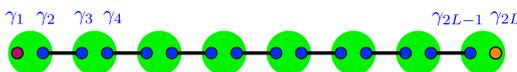
>
> One finds unpaired Majorana modes $\gamma_1$ and $\gamma_{2L}$.
>
> (IV) According to II, one can form a non-local physical operator, $d^\dagger = (\gamma_1 - i\gamma_{2L})/2$, associated with a state which is either empty ($|0\rangle$) or occupied ($|1\rangle$), the Majorana bound state. The former corresponds to the Bogoliubov vacuum and, formally, the latter is an excitation but with zero energy: both states are degenerate. They form a topological quantum bit.
>
> (V) Particle number is not conserved in a superconductor, but fermion parity is: $|0\rangle$ has even parity, and $|1\rangle$ has odd parity. These states differ by one charge, and are not practically useful as a qubit.
> As discussed in Sec. 7, the solution is to add one extra pair of Majorana modes (i.e., an additional chain) to form one logical qubit. In general, M+1 pairs of Majorana modes (dense quantum bit encoding) constitute M logical quantum bits; alternatively, 2M pairs of Majorana modes (sparse quantum bit encoding) constitute M logical quantum bits.

energy $\Delta_p$, a topological transition can occur as a function of the relative strength of the nearest-neighbor hopping parameter *t* and the normal-state chemical potential or on-site energy $\mu$. The topologically non-trivial phase is characterized by a topological gap $\Delta_{\text{topo}}$ in the band structure which is usually smaller than the superconductor gap energy, $\Delta_{\text{topo}} < \Delta_p$. This can be true, however, also for a topologically trivial superconducting phase, $\Delta_{\text{triv}} < \Delta_p$. Unlike a trivial phase, the topological phase is accompanied by the emergence of zero-energy modes at both ends of the chain, the so-called Majorana bound state. Box 1 explains the physics of the Kitaev chain and its Majorana bound state. By weighting the electron-contribution (in contrast to the hole-contribution) the superconducting band structure reveals some characteristic *inversion* of spectral weight in the topological phase (Fig. 1(A)); this inversion is also visible in real space in the local density of states (LDOS) as a function of energy *E* and position *x*, LDOS(*E*, *x*) (Fig. 1(B)).

To reveal the connection between atomic spin chains on superconducting substrates and the Kitaev chain model, it is instructive to first consider a single magnetic impurity on a (spin–orbit coupled) superconducting substrate within the energy gap of the SC substrate [64–67]. These discrete states result from the Cooper-pair breaking effect of magnetic impurities interacting with the SC [64–70]; for details see Box 2, which contains Refs.[50–52,64–86]. A second magnetic impurity adjacent to the first one leads to more energy states within the SC gap and so on, until states from multiple impurities hybridize and form bands which may or may not be topologically non-trivial (see Box 3 for further details;

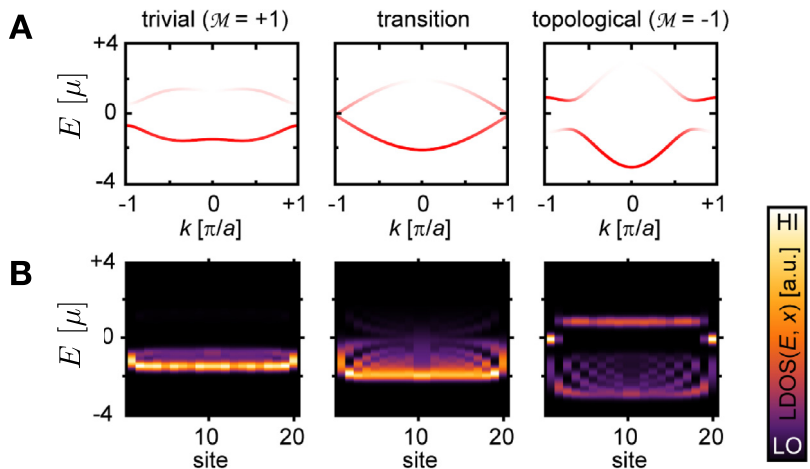

**Fig. 1.** Kitaev model for a 1D topological superconductor. (A) Topological phase transition in the Kitaev chain: the band structure changes from a trivial regime (left) through a gap-closing (middle) to a topologically non-trivial regime (right) with increasing nearest-neighbor hopping $t$. (B) The corresponding LDOS$(E, x)$ plots for a chain with 20 atomic sites. In the topologically non-trivial regime (right), zero-energy Majorana modes emerge at both chain ends. They always appear as pairs and are robust against reasonably small perturbations. Parameters used: $t = 0.2\,\mu$ (left), $t = 0.5\,\mu$ (middle), and $t = 1.0\,\mu$ (right).

---

**Box 2: Yu-Shiba-Rusinov states**

A magnetic impurity in a conventional s-wave superconductor induces subgap states at energies $\pm E_J$ within the gap of the superconductor (cf. DOS plot in Box 3). These localized states are referred to as Yu-Shiba-Rusinov (YSR) states [64–66]. The basic idea is that the magnetic impurity violates time-reversal invariance and locally suppresses the superconducting order parameter. Assuming classical spins $S$ with magnetic moment $J$, the YSR bound state energies are given by

$$E_J = \frac{1-\alpha^2}{1+\alpha^2}$$

with $\alpha = JS\pi\nu_0/2$ and the density of states $\nu_0$. Due to particle-hole symmetry the YSR states must come in pairs (positive and negative energy) but their spectral weight can be asymmetric. The YSR states are spin-polarized. For details we refer to the extensive review article by Balatski et al. [67].

Using STM techniques, the first successful measurement of YSR states was reported for a Mn atom on a superconducting Nb(110) surface [68]. Excitations over a few atomic diameters around the impurity at the surface could be detected. Significant improvements of the energy resolution could be achieved by using superconducting STS tips [69]. There has been notable progress on experimental STM work on YSR states over the past couple of years [70–78]. In particular, it has been demonstrated that the energies of YSR states of an individual magnetic adsorbate on a superconductor can critically depend on its adsorption site relative to the substrate lattice [52, 79]. Therefore, it is decisive that all magnetic adatoms within a YSR chain occupy the same type of adsorption site in order to lead to a coherent quantum state.

The spatial extension of YSR states was in the focus of experimental STM/STS studies in Refs. [80, 81]. Long-range YSR states are favorable in view of the hybridization between YSR states of individual magnetic impurities over large distances. The investigation of the orbital character of YSR states was first reported in Refs. [82, 83]. It plays a crucial role for the degree of hybridization of YSR states in spin chains along different crystallographic directions with respect to the superconducting substrate lattice (see Figure 8).

Spin-polarized STM/STS measurements of YSR states of individual magnetic adsorbates on superconducting substrates were first achieved in Ref. [84] allowing for an even deeper insight into the nature of sub-gap states [85] as well as a more involved comparison with theory. In the context of YSR chains, spin-polarized STS measurements are key as they allow for the distinction between Majorana zero modes and trivial states at or close to zero energy [50–52], in line with theoretical work [86].

note that the oscillations at larger energies of the DOS plots in Box 3 are due to finite size).

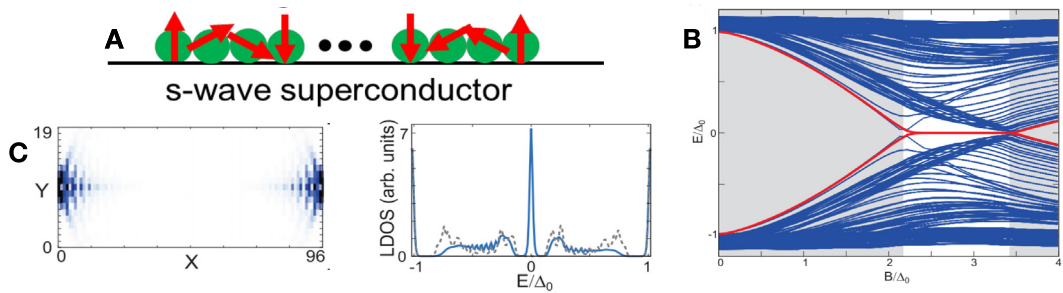

**Fig. 2.** Theoretical prediction of Majorana quasiparticles in an atomic-scale magnet–superconductor hybrid system. (A) If an atomic spin chain with a helical magnetic state is proximitized to an s-wave superconductor, a topological superconducting state can emerge similar to the Kitaev chain. (B) Theoretically predicted band gap closing as a function of magnetic exchange field strength. The zero-energy modes, which emerge for a particular regime of exchange field, should be localized at both ends of the chain and can be detected by observing a characteristic zero-energy peak in the local density of states (LDOS), e.g. by scanning tunneling spectroscopy (STS). (C) Majorana modes always appear in pairs at both ends of spin chains proximitized to a superconducting substrate with s-wave pairing.
*Source:* Taken from Ref. [11].

As a consequence, an atomic spin chain proximity-coupled to an s-wave superconductor constitutes an experimental realization of the Kitaev model [3]. This can be theoretically further substantiated by projecting the spin chain–superconductor Hamiltonian onto the lowest band, leading to an effective one-band Kitaev model, potentially with longer-ranged hoppings and pairings [7,8,13]. It has been theoretically predicted first that a spin chain with a helical magnetic texture [7,8,11–16,23,87], which is proximitized to an s-wave superconductor (Fig. 2(A)), can enter a topologically non-trivial superconducting state with a gapped band structure in the interior ("bulk") of the chain and Majorana zero-energy modes (MZM) at the chain's ends ("boundaries"). In order to stabilize the helical spin state, a sufficiently strong Rashba spin–orbit coupling, giving rise to significant Dzyaloshinskii–Moriya interactions [88,89], is required [16]. Moreover, it was shown that the effective magnetic exchange field, i.e., the effective magnetic moment, needs to be tuned in order to enter the topologically non-trivial regime [11] (Fig. 2(B)). That is, it might require a careful choice of magnetic adatoms to hit the topological phase. The strong localization of the MZMs in the early experiments of Ref. [46] appeared to be surprising as one would naively expect that the Majorana localization length $\xi_M = \hbar v_F / \Delta_{\text{topo}}$ is much larger than the superconducting coherence length $\xi = \hbar v_F / \Delta$ with superconducting order parameter $\Delta$. An explanation was given in terms of a renormalized quasiparticle weight $Z$ which is caused by the coupling of the adatoms to the superconductor; as a consequence, the Fermi velocity is renormalized, $v_F \to \tilde{v}_F = Z v_F$ which can lead to a Majorana localization length of the order of the distance between adatoms [20].

Such highly localized features can be spatially resolved by high-resolution scanning tunneling microscopy (STM) techniques [44]. In addition, the characteristic zero-energy peak in the LDOS, as shown in Fig. 2(B), can be resolved by scanning tunneling spectroscopy (STS) at sufficiently low temperatures. Importantly, the MZM always show up in pairs at both chain ends (Fig. 2(C)), in contrast to zero-bias states of trivial origin.

Besides helical spin chains, also ferromagnetic (FM) atomic chains proximity-coupled to an s-wave superconductor can lead to non-trivial topological phases and associated MZM, as predicted by subsequent theoretical studies [17,19–22] (Fig. 2(C)). It has been shown theoretically that the system with helical spin texture is equivalent to the one with a ferromagnetic structure with an anisotropic spin–orbit coupling and shifted chemical potential [8,90–92]. Both systems thus belong to the same universality class. It is instructive to consider the effects of the different contributions of a ferromagnetic spin chain on a spin–orbit coupled superconducting substrate: the spin–orbit coupling is essential to break the spin-symmetry and mix the different angular-momentum contributions, while the effective magnetic field breaks time-reversal symmetry and gaps out part of the normal-state band structure. The induced superconductivity thus acts on a non-degenerate Fermi surface (assuming the chemical potential is properly chosen) and results in a topologically non-trivial superconductor (see Box 3 for details).

Even Shiba chains with collinear antiferromagnetic (AFM) spin order can show non-trivial topological phases and associated MZMs [93,94]. Interestingly, it was found that in general the topological gap is larger for AFM chains than for FM ones, yielding a more pronounced localization of MZMs in AFM chains. For a detailed comparison between theory and experiment, the determination of the magnetic texture of the atomic chains is mandatory because it largely influences the topological phase diagram for such chain systems. This requires the application of spin-resolving techniques with atomic-scale spatial resolution, such as SP-STM [44].

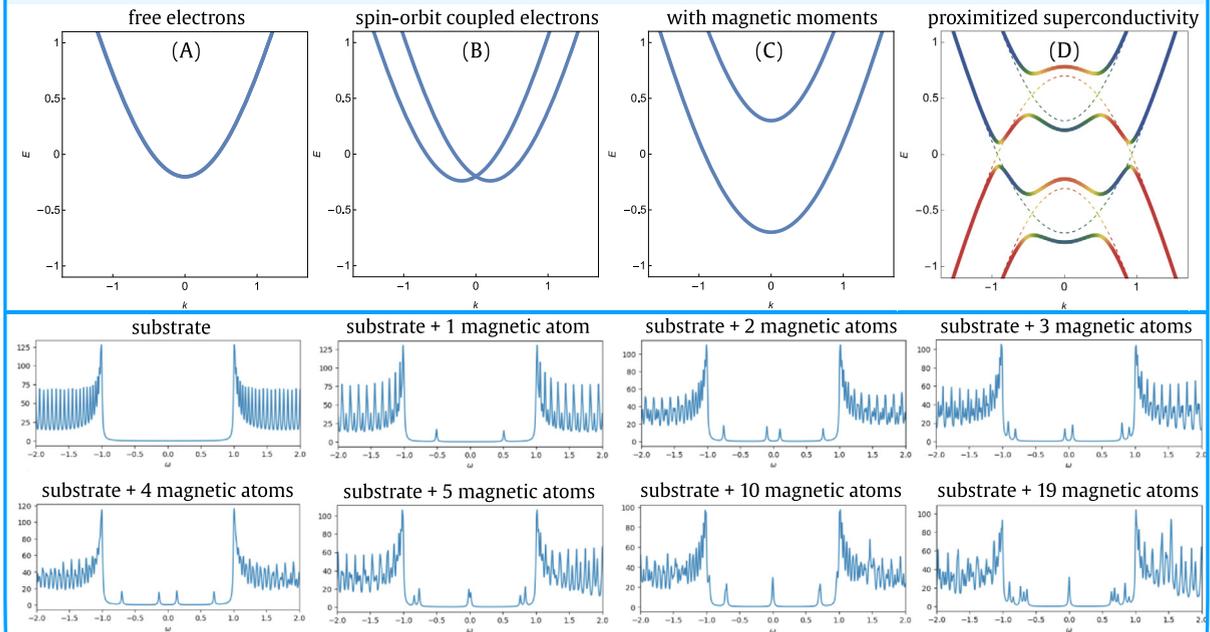

Box 3: Yu-Shiba-Rusinov chain - momentum space vs. real space picture

Proximity-induced topological superconductivity, here illustrated for a magnet-superconductor hybrid chain. (A) Free electrons with spin-degenerate bands. (B) Splitting of and spin-mixing within these bands due to Rashba spin-orbit coupling. (C) Magnetic moments induce a Zeeman splitting and open a gap at $k = 0$. (D) Proximity-induced superconductivity applied to the non-degenerate bands at the Fermi energy results in a topological superconducting regime, as indicated by the partial inversion of particle- (blue) and hole-weight (red). Dashed lines in D correspond to the electron-hole "doubled" bands from C.

Instead of the momentum space perspective (i.e., infinitely large system), one can consider the superconducting substrate involving spin-orbit coupling and build the chains atom-by-atom. The substrate here is a chain consisting of 99 atoms, with superconducting coherence peaks at energies ω = ±1 (arbitrary units). Magnetic atoms are placed in the middle of the substrate. Already for 10 atoms, the Majorana peak at ω = 0 is clearly visible.

The first experimental signatures of MZM were found for self-assembled FM Fe chains on superconducting Pb(110) substrates [46,48,49,51]. STS spectra revealed a zero-bias conductance peak at one end of the self-assembled Fe chains being absent in the interior of the chains [46,48]. A more complex spatial pattern of the zero-energy mode in terms of a 'double-eye' feature was found at higher spatial resolution [49]. Moreover, the spin-dependent characteristics of the zero-energy end states were probed by SP-STM [51]. However, subsequent theoretical studies showed that the observation of a zero-energy peak at one chain end by STS is not sufficient for claiming the existence of Majorana quasiparticles [21,22], but can also arise for topologically trivial magnetic chains interacting with a superconducting substrate. Several important additional tests for establishing the Majorana nature of the observed spectroscopic features could not be performed based on the self-assembled Fe chains due to atomic intermixing with the Pb substrate as a result of the annealing procedure required for the self-assembly process. The Fe chains were not atomically well defined but contained clusters and impurities of unknown origin, leading to uncontrollable disorder on the atomic scale. Consequently, it could not be tested whether the zero-energy modes appear simultaneously at both chain ends (Fig. 2(C)) and whether they form a quantum-coherent state over the whole chain length. In addition, the atomic-scale disorder as present in the self-assembled magnetic chains leads to numerous low-energy states in STS spectra. This might be the reason why a topological gap could not be resolved in these early experiments, even when the STS measurements were performed at sub-Kelvin temperatures [49]. It was concluded [49] that more well defined and simpler chains should be built by making use of STM-based single-atom manipulation techniques.

Furthermore, the self-assembly approach does not allow for a systematic study of the emergence of Majorana bound states as a function of chain length which, however, would be desirable for a detailed comparison with theory. Moreover, the geometrical arrangements of magnetic adatoms on single-crystal substrates, which can be realized based on self-assembly processes, are rather limited (e.g. linear adatom chains can be achieved on two-fold symmetric crystal lattice substrates such as the (110) surfaces). However, in order to prepare more complex geometries, such as T-junctions, Y-junctions, rings, or complex 2D networks and circuits, as proposed by numerous theory groups [30,60,62,63], a more versatile technique for the fabrication of suitable atomic-scale platforms for topological quantum computing needs to be developed.

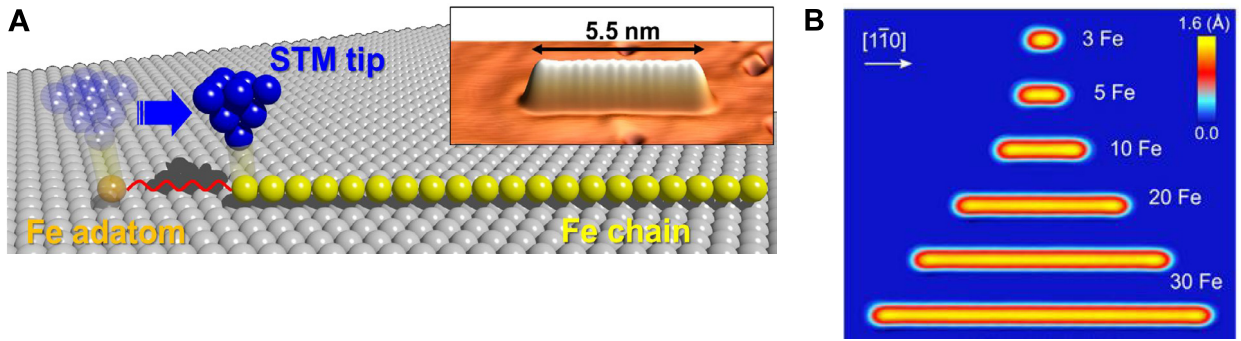

**Fig. 3.** Bottom-up fabrication of atomic spin chains on a superconducting substrate. (A) Schematic of the atom-by-atom construction of a magnetic chain of Fe atoms on a Re(0001) single-crystal substrate by STM-based single-atom manipulation. (B) STM image of artificially constructed atomic Fe chains of variable length (3 up to 40 Fe atoms) on a superconducting Re(0001) substrate.
*Source:* Taken from Ref. [52].

## 3. Fabrication of disorder-free atomic spin chains on clean superconducting substrates

The design, control, and manipulation of Majorana bound states based on magnetic atomic chains on superconducting substrates require well-defined defect-free model-type structures fabricated with atomic-scale precision. The method of choice for achieving this goal is STM-based single-atom manipulation which has first been demonstrated in the early nineties by D. Eigler and coworkers [95,96]. It was shown that diluted adatoms on surfaces deposited from the gas phase onto an atomically clean substrate can be manipulated in various ways by an atomically sharp STM tip.

Lateral adatom manipulation on the substrate surface can be achieved by making use of either pushing, pulling, or sliding modes of operation; the STM tip has to be lowered towards the adatom by reducing the tunneling gap resistance compared to the STM imaging mode and subsequently brought to a new designated adatom position by means of piezoelectric drives; forces acting between the front tip-atom and the surface adatom have to be controlled such that the adatom can be moved laterally over the surface, overcoming the potential barriers between favorable adsorption sites, but preventing a transfer of the surface adatom towards the STM-tip.

Vertical adatom manipulation involves a pick-up of the surface adatom by the STM-tip induced, e.g. by a voltage pulse applied between tip and sample. Subsequently, the STM-tip is brought to the new designated adatom position and a voltage pulse of opposite polarity is applied to transfer the adatom from the tip back to the substrate surface. The vertical adatom manipulation mode is usually applied for adatoms on metal substrates covered with an ultrathin oxide or nitride layer. For a successful atomic-scale fabrication of model-type structures consisting of magnetic atomic chains and more complex adatom arrangements on conventional superconducting substrates, the STM-based single-atom manipulation methods have to work on various surfaces of such conventional superconductors. While lateral adatom manipulation was first established for adatoms on non-superconducting metal substrates (e.g. Ni(111), Cu(111), Ag(111), Pt(111)), it became possible in recent years to fabricate reasonably long magnetic adatom chains on various surfaces of elemental superconductors, including Re(0001) [52,54], Nb(110) [55,57–59] and Ta(110) [97]. In a few cases, superconducting alloys, such as $\beta$-Bi$_2$Pd, were chosen as substrates for the bottom-up assembly of atomic spin chains [56].

An example of an artificially created, defect-free atomic Fe chain on Re(0001) [52] is shown in Fig. 3(A). Such chains are assembled atom-by-atom so that any chain of length between a single adatom and the finally achieved chain length can be studied as well (see Fig. 3(B)). The maximum chain length achievable is limited by the residual defect density of the single-crystal substrates and typically varies between several ten and several hundred adatoms [54]. Each adatom being added to such chains is previously characterized spectroscopically by STS to make sure that only adatoms of the intended species are part of the artificially constructed chains. Therefore, the presence of any kind of disorder – even a single impurity – in the chain can be excluded.

The atomic spin chains can be built along different crystallographic directions with respect to the substrate lattice. It is usually possible to realize close-packed adatom chains or diluted adatom chains, e.g. with a spacing of two or three substrate lattice periods. Adatoms residing on inequivalent adsorption sites (e.g. fcc- or hcp-sites) usually have different electronic and magnetic properties (e.g. different magnetic moments and magnetic anisotropies), therefore the control of the adsorption site of each individual atom during the construction of the artificial chains is decisive. The size of the magnetic moments and magnetic anisotropies can also be tuned by the choice of the magnetic adatom species (e.g. Cr, Mn, Fe, Co) and the choice of the substrate. The chosen combination of adatom species and substrate also determines the strength of Dzyaloshinskii–Moriya-type interactions [98,99] and the likelihood that a spin spiral ground state of the magnetic adatom chain can be achieved or not. The electronic and magnetic coupling between the adatoms and the substrate is further tunable by introducing atomically thin oxide or nitride films on the superconductor's surface. Examples are artificially built atomic spin chains made of Fe adatoms on a Ta(001)-O substrate [53]. In this case, STM-based vertical atom manipulation techniques need to be employed since lateral adatom manipulation hardly works on oxide covered surfaces.

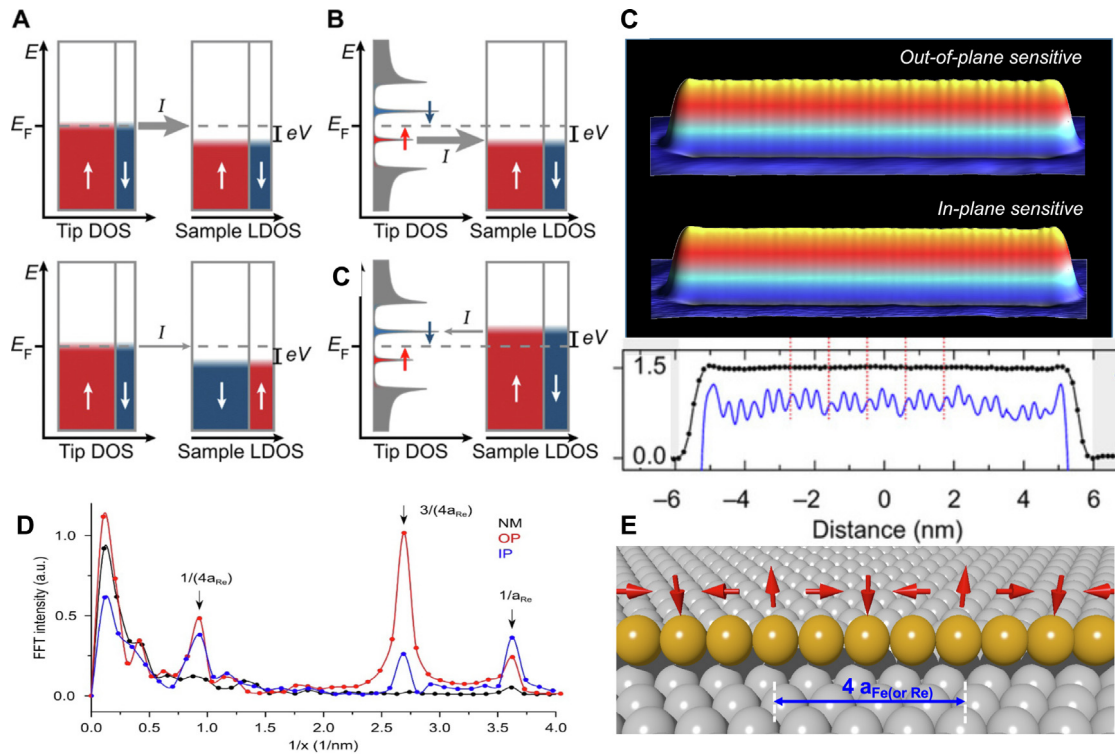

**Fig. 4.** Magnetic characterization of atomic spin chains on superconducting substrates. (A) Concept of spin-polarized tunneling involving a magnetic STM tip. The magnitude of the spin contrast is determined by the imbalance of spin-minority and spin-majority states of both sample and STM tip. (B) Concept of spin-polarized tunneling involving 100% spin-polarized discrete (Shiba) states of a magnetic atom at the front end of a superconducting STM tip, thereby maximizing the spin contrast in SP-STM imaging. [Taken from Ref. [45]]. (C) SP-STM images of a bottom-up fabricated atomic Fe chain on Re(0001) revealing the out-of-plane spin component (top) and the in-plane spin component (bottom) of the non-collinear spin spiral ground state. Bottom: Constant-current STM profile obtained with a non-magnetic tip (black, vertical axis given in Å) and corresponding magnetic profile (blue) obtained from the SP-STM image measured with the out-of-plane spin-sensitive tip. (D) 1D-Fast Fourier Transform (FFT) of the STM- and SP-STM line profiles obtained with non-magnetic (black), out-of-plane spin-sensitive (red), and in-plane spin-sensitive (blue) STM tips. With both types of spin-sensitive tips, a superstructure with a period of four times the atomic lattice constant is revealed, thereby proving a non-collinear spin spiral state. (E) Schematic of the spin spiral with a 4-atom periodicity along the Fe chain as deduced from the experimental SP-STM data shown in (C) and the FFT analysis in (D).
*Source:* Taken from Ref. [52].

## 4. Revealing the spin structure of atomic chains by spin-polarized scanning tunneling microscopy

After determining the structural properties of the atomic chains on superconducting substrates by STM-based bottom-up fabrication techniques, as described in the previous chapter, it is decisive to characterize their magnetic ground states. Even for simple models, the topological phase diagrams are usually different for ferromagnetic and antiferromagnetic orders or for magnetic moments with out-of-plane and for those with in-plane orientation. Although a spin-spiral can be mapped to a ferromagnetic structure [92] the phase diagrams might still deviate due to the shifts of involved parameters. For realistic multi-orbital magnetic chain–superconductor hybrid systems as studied experimentally, the influence of the magnetic order on the topological phase diagram and spectral gap size can be even more drastic.

Over the past decades, SP-STM [44,100] has developed into a powerful technique for revealing complex 3D spin textures with atomic-scale spatial resolution [88,101–104]. The magnitude of the spin contrast in spin-polarized tunneling experiments is determined by the imbalance of spin-minority and spin-majority states of both electrodes, i.e. sample and STM tip (Fig. 4(A) and (B)). Importantly, SP-STM is sensitive to the projection of the local surface magnetization (or the magnetic moment of individual adatoms) onto the quantization axis given by the spin orientation of the front tip-atom [44], therefore it is extremely important to be able to control the spin orientation of the front atom of the SP-STM tip. This can be achieved by at least two different experimental strategies.

By orientating the spin of the front tip-atom by an externally applied 3D vector field [102], all three components of the surface spin distribution can be mapped down to the atomic scale [44]. However, this experimental approach is only reasonable if one can exclude a simultaneous response of the sample under investigation to the external B-field, otherwise the interpretation of the experimental data might become difficult. Alternatively, SP-STM probe tips with an antiferromagnetic thin film coating, where the spin orientation of the front end of the tip can be controlled by the intrinsic thin film magnetic anisotropy [44,101], can be used for mapping different surface spin components. For instance, ultrathin Cr-coatings on W-tips lead to an out-of-plane magnetic sensitivity, while thicker Cr-coatings lead to an in-plane magnetic sensitivity [101]. This behavior is caused by a thickness-dependent reorientation transition of magnetic anisotropy in the ultrathin Cr-films covering the W-probe tips. However, the in-plane spin orientation direction of Cr-coated tips has to be calibrated on well-defined 3D-spin textures. This can easily be done by preparing larger 2D islands of magnetic transition metals on substrates such as Re or Ta, which usually exhibit complex 3D spin textures due to sizable interfacial Dzyaloshinskii–Moriya interactions. For example, monolayer islands of Fe on Re(0001) exhibit both a 120° Néel-ordered state as well as a spin-spiral state in the vicinity of dislocation lines [105,106]. Both types of 3D spin textures are ideally suited to calibrate the direction of spin orientation at the front end of SP-STM tips.

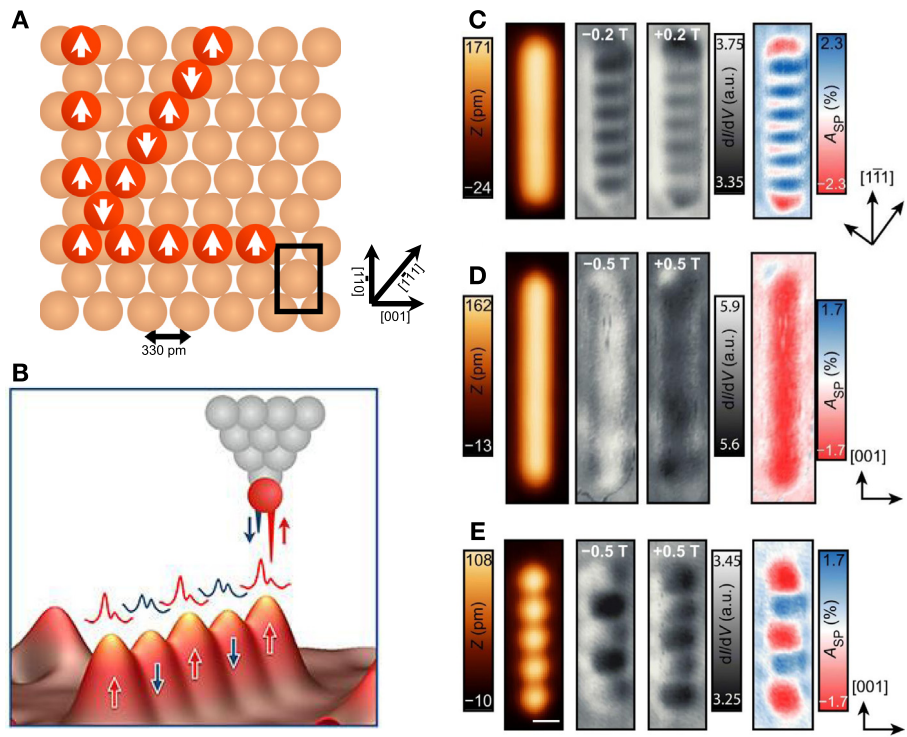

**Fig. 5.** Magnetic state of atomic spin chains as a function of crystallographic orientation. (A) Schematic of the magnetic ground states of atomic Fe chains on Nb(110) in different crystallographic directions. (B) A Shiba-state SP-STM tip is used to probe the antiferromagnetic ground state of a chain of Fe atoms on a Nb(110) substrate. (C) Antiferromagnetic ground state of a closed-packed linear chain of Fe atoms on Nb(110) in [1$\bar{1}$1]-direction as revealed by SP-STM. (D) Ferromagnetic ground state of a closed-packed linear chain of Fe atoms on Nb(110) in [001]-direction as revealed by SP-STM. (E) Antiferromagnetic ground state of a linear chain of Fe atoms on Nb(110) in [001]-direction with a doubled inter-atomic spacing as compared to (D).
*Source:* Taken from Ref. [45].

Examples of SP-STM images of a bottom-up fabricated 40-atom Fe chain on Re(0001) are presented in Fig. 4(C). The Fe atoms are close-packed along the chain and they all reside on hcp adsorption sites with respect to the Re(0001) substrate lattice. Both the out-of-plane spin component (top) and the in-plane spin component (bottom) of the magnetic texture present in the Fe chain have been mapped using different spin-sensitive STM probe tips [52]. A magnetic superstructure with a period of four atomic lattice spacings is clearly visible for both spin channels, thereby indicating a non-collinear spin spiral ground state. Fig. 4(E) presents a schematic of the spin spiral with a 4-atom periodicity along the Fe chain as deduced from the experimental SP-STM data sets shown in Fig. 4(C) and (D).

It should be emphasized that a detailed characterization of the evolution of the magnetic texture as a function of chain length requires SP-STM tips being capable of both single-atom manipulation and spin-sensitive imaging. Since 2010 it has become possible to perform single adatom manipulation with an atomically sharp spin-sensitive STM tip while preserving the spin orientation of the front tip-atom during each step of the manipulation process [107–109]. This way, one can switch multiple times between adatom manipulation and subsequent spin-state imaging by SP-STM without changing the spin contrast [52] (e.g. by a spin reorientation of the front tip-atom).

In contrast to self-assembly processes, STM-based single-atom manipulation allows the fabrication of atomic chains along different crystallographic directions with respect to the substrate lattice. Subsequent SP-STM studies can then reveal the dependence of the magnetic ground states on the chain direction. An example is presented in Fig. 5 where different types of atomic Mn chains on a Nb(110) substrate are considered. It was found [45] that close-packed Mn chains in [1$\bar{1}$1]-direction exhibit an AFM ground state (see schematic in Fig. 5(A) and experimental SP-STM data in Fig. 5(C)). In order to enhance the spin contrast for close-packed atomic chains, a novel SP-STM method based on the use of a Shiba-state tip was introduced [45] (see Fig. 5(B)).

The electron- and hole-like pairs of YSR-states are 100% spin-polarized, as experimentally confirmed by SP-STM [84], and therefore can be employed as a most efficient spin-sensitive local probe [45]. Starting with superconducting STM-tips made of Nb, Ta, or Re, which are functionalized by an individual magnetic atom or small magnetic cluster at the tip apex (Fig. 5(B)), high spin contrast imaging in SP-STM can routinely be achieved [45]. In contrast to close-packed atomic Mn chains on Nb(110), running along the [1$\bar{1}$1]-direction, the close-packed Mn chains built along the [1$\bar{1}$0]- and [001]-directions with respect to the Nb(110) substrate lattice exhibit FM ground states [45] (see schematic in Fig. 5(A) and experimental SP-STM data for the [001]-chain in Fig. 5(D)). If the spacing between the Mn adatoms is increased, different magnetic ground states can emerge. For instance, an atomic Mn chain in [001]-direction with a double spacing with respect to the atomic lattice of the Nb(110) substrate exhibits an AFM ground state [45] (Fig. 5(E)). In general, the magnetic ground state of transition metal chains on superconducting substrates depends critically on the crystallographic direction with respect to the substrate lattice as well as on the interatomic distances within the chains. Furthermore, the magnetic ground state of transition metal chains (collinear FM or antiferromagnetic vs. non-collinear spin spirals) can be largely influenced by the choice of the superconducting substrate. Heavy elements, like Re with sizable Dzyaloshinskii–Moriya interactions, favor non-collinear spin spiral states whereas lighter elements, like Nb, are likely to favor collinear (FM or AFM) ground states.

In conclusion, the unambiguous determination of the magnetic texture within atomic spin chains on superconducting substrates by SP-STM methods is decisive in view of the emergence of non-trivial topological phases in such magnet–superconductor hybrid systems, in particular for a detailed comparison between theoretical predictions and experimental results [94,110].

## 5. Evolution of sub-gap Shiba states from individual magnetic adatoms to atomic spin chains on s-wave superconductors

In this chapter, we will focus on experiments showing the evolution of sub-gap Shiba states from individual magnetic adatoms up to extended linear atomic chains on various types of superconducting substrates. Since the energy gap of conventional elemental superconductors is small (on the order of 1 meV), experimental studies of sub-gap states require extremely high energy resolution (on the order of several micro-eV). This can only be achieved by operating the STM instrument at ultra-low temperatures (well below 1 K) and by making use of superconducting STM-tips. Superconducting tips (made of, e.g., Nb, Ta, or Re) lead to S-I-S tunnel junctions (where S stands for superconductor and I for insulator; the latter is in fact a vacuum barrier), thereby allowing for a considerable improvement of the energy resolution, as demonstrated previously for STS studies of YSR states of individual magnetic adatoms or molecules on superconducting substrates [52,69,111,112]. Importantly, the STM-based single-atom manipulation methods, as described in chapter III, have to work with superconducting probe tips as well, so that atomic-scale fabrication and high-resolution spectroscopic characterization can directly be combined in a single experiment. This is mandatory if one would like to study, for instance, the emergence of zero-energy Majorana bound states as a function of the number of adatoms within the chain, i.e. the chain length.

As a first example for such type of experiments, the evolution of YSR states of individual Fe adatoms up to linear chains of several ten Fe adatoms on superconducting Re(0001) have been studied by STS at 350 mK [52]. It was found that Fe atoms adsorbed on fcc sites with respect to the Re(0001) lattice exhibit a YSR bound state energy of 0.1 meV, whereas Fe atoms adsorbed on hcp sites show a YSR bound state energy of 0.02 meV only. This means that YSR states are very sensitive to the type of adsorption site which need to be determined by STM with atomic spatial resolution. By approaching two Fe adatoms on hcp-adsorption sites to an interatomic distance of 0.274 nm (corresponding to the lattice period of Re(0001)), the YSR-related peaks in the tunneling spectrum are shifted to 0.11 meV, indicating that the Fe adatoms are interacting with each other and hybridization between YSR states occurs [52]. As further Fe adatoms on hcp adsorption sites are added to form close-packed Fe chains of increasing length (see Fig. 3(B)), hybridized Shiba bands emerge within the energy gap ($\Delta_{Re}$ = 0.245 meV) of the superconducting Re(0001) substrate (see Fig. 6(A) for a 40-atom long Fe chain). Besides the Shiba states showing up at finite energies and in the interior of the Fe chain, a significantly enhanced spectroscopic feature at zero energy appears at both ends of the defect-free 40-atom Fe chain on superconducting Re(0001) [52], as can be seen in the corresponding panel of Fig. 6(A). The increased spectral weight at zero energy at both ends of the Fe chain can also be seen in the corresponding line profile of Fig. 6(B) which additionally shows the profiles of the YSR-states at +0.12 meV and −0.12 meV. For comparison, a line profile of the spectroscopic data measured at an energy of −0.65 meV, i.e. outside of the superconductor's energy gap, is presented showing the absence of increased intensity at both chain ends. For reference, the topographic line profile as well as the spin-resolved SP-STM profile of the Fe chain's spin spiral state are presented at the bottom of Fig. 6(B).

In order to study the evolution of the zero-energy spectroscopic features as a function of chain length, detailed STS characterization has been performed for all chain lengths, from the single Fe adatom up to the 40 Fe-atom long chain [52]. The corresponding zero-bias conductance (ZBC) profiles for some chain lengths are presented in Fig. 6(C), showing a gradual increase of spectral weight at both chain ends as the Fe chains get longer. The experimentally observed behavior can well be reproduced by theory. Fig. 6(D) shows the theoretically predicted behavior of the zero-energy states for atomic Fe chains of various length on superconducting Re(0001). For short chain lengths (below 20 Fe atoms), there is a significant hybridization of the zero-energy states of both ends observable. However, as the Fe chain gets longer, the zero-energy states of both chain ends become increasingly separated.

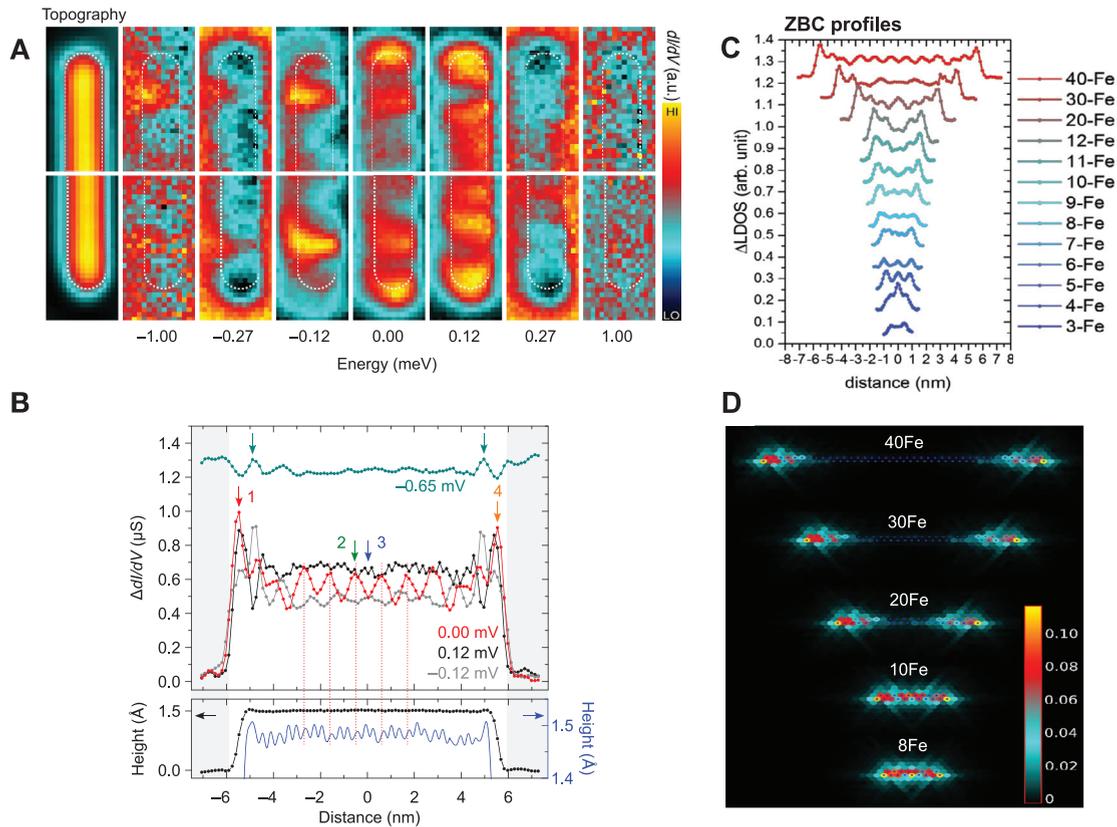

**Fig. 6.** Emergence of zero-energy modes at both ends of bottom-up constructed atomic Fe chains on Re(0001) as a function of chain length. (A) Spatial maps of the topography (left) of a 40-atom Fe chain on superconducting Re(0001) together with a series of differential tunneling conductance maps at different energies given in meV. A significant spectral weight at both ends of the Fe chain on superconducting Re(0001) appears at zero energy. (B) Line profiles of the topographic height and spin spiral texture (bottom) as well as of the differential tunneling conductance as measured by STS at 350 mK, reflecting the LDOS distribution at different energies (top). (C) Zero-bias conductance (ZBC) profiles along the Fe chains of variable length revealing an enhancement at both ends for increasing chain length. (D) Theoretically calculated zero-energy modes for atomic Fe chains of variable length on Re(0001) showing strongly hybridized states for short chain lengths and increasingly separated zero-energy modes for larger chain lengths.
*Source:* Taken from Ref. [52].

The topological phase diagram for this magnet–superconductor hybrid system as a function of chemical potential and exchange interaction parameter is presented in Fig. 7(A). For the ab-initio derived values of chemical potential and exchange parameter (see yellow arrow), a topologically non-trivial state of the Fe-Re hybrid system is predicted [52]. This means, that the zero-energy modes as experimentally observed for this particular hybrid system can be assigned to Majorana modes. However, due to the limited STS energy resolution at the measurement temperature of 350 mK, these zero-energy Majorana states overlap with finite-energy Shiba states. This conclusion can be supported by further theoretical analysis taking temperature effects into account. Fig. 7(B) shows the theoretically derived Shiba bands in a 20-atom Fe chain on superconducting Re(0001) taking thermal broadening at a temperature of 350 mK into account. At both ends of the chain, the finite-energy Shiba states overlap with the zero-energy modes being expected for a topologically non-trivial hybrid system. However, by reducing the thermal broadening by a factor of ten in the calculation, the zero-energy Majorana modes appear within a clearly visible topological gap separating them well from the finite-energy Shiba states [113].

As a consequence of this interesting theoretical insight, the measurement temperature for the magnetic Fe chain on superconducting Re(0001) hybrid system should be lowered at least by a factor of ten. Alternatively, one can choose a superconducting substrate with a larger superconductor-normal metal transition temperature $T_c$ and therefore a larger superconductor gap energy, thereby facilitating the observation of sub-gap states considerably. Before discussing alternative substrates, such as Nb or Ta, it should be noted that the signatures of Majorana zero-energy modes, as found for close-packed atomic Fe chains with hcp adsorption sites on superconducting Re(0001), could not be found for Fe chains with fcc adsorption sites, nor for Fe chains with a doubled interatomic spacing, i.e. for dilute Fe chains. Furthermore, other types of transition metal chains on Re(0001), e.g. made of Mn or Co, did not show indications for the existence of Majorana states either [54]. This can be understood by a systematic trend of the YSR states of Mn, Fe, and Co adatoms on superconducting Re(0001), which enables the identification of the most promising hybrid system for the realization of topological superconductivity [79]. This is indeed the system of hcp Fe adatom chains on superconducting Re(0001). In addition to the near zero-energy YSR states of hcp Fe adatoms on superconducting Re(0001), the preference of a non-collinear spin spiral ground state, as the close-packed Fe chains are built up atom-by-atom on the Re(0001) substrate, is expected to additionally favor the emergence of a topologically non-trivial phase and associated Majorana states.

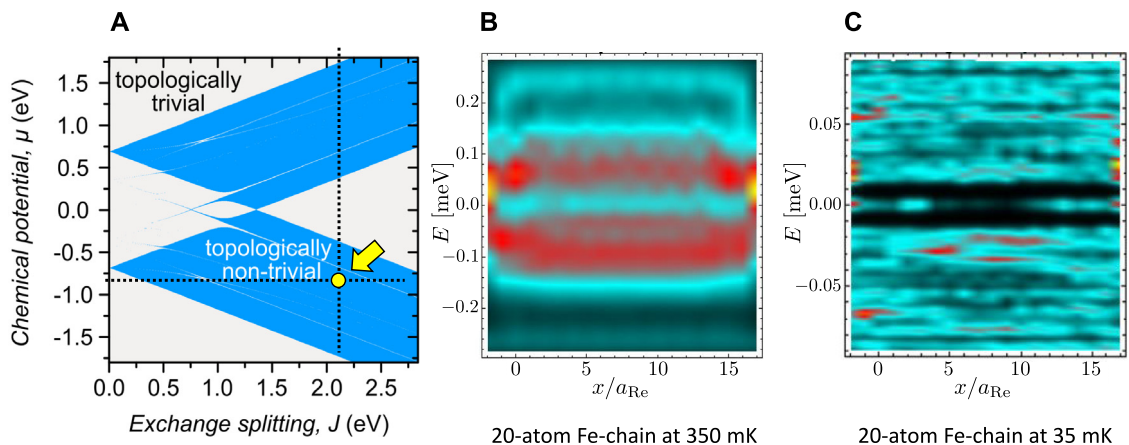

**Fig. 7.** Theoretical results for close-packed atomic Fe chains on superconducting Re(0001). (A) Topological phase diagram as a function of chemical potential and exchange interaction parameter. For the ab-initio derived values of chemical potential and exchange parameter, a topologically non-trivial state of the Fe-Re hybrid system is predicted. [Taken from Ref. [52]]. (B) Theoretically derived Shiba bands in a 20-atom Fe chain on superconducting Re(0001) taking thermal broadening at a temperature of 350 mK into account. At both ends of the chain, the finite-energy Shiba states obviously overlap with the zero-energy modes being expected for a topologically non-trivial hybrid system. (C) By reducing the thermal broadening by a factor of 10 in the calculation, the zero-energy modes within a clearly visible topological gap become energetically well separated from the finite-energy Shiba states.
*Source:* courtesy of Th. Posske.

Niobium substrates appear to be the optimum choice for realizing magnet–superconductor hybrid systems because the gap energy of niobium is largest among all elemental superconductors and therefore the observation of sub-gap states arising from the interaction with magnetic adatoms is facilitated. However, the surfaces of niobium single crystals are difficult to clean using conventional surface science techniques. Only recently, sufficiently clean niobium substrates for well defined experiments on atomic-scale magnet–superconductor hybrid systems could be prepared [45,55,114–116]. Among the magnetic transition metal elements, Mn was found to induce sub-gap YSR states on Nb(110) substrates which are most promising in view of the possible emergence of topologically non-trivial states of Mn chains on superconducting Nb.

Fig. 8 shows the evolution from discrete Shiba states of a single Mn adatom to Shiba bands for long atomic Mn chains on superconducting Nb(110) being assembled atom-by-atom with the STM. An individual Mn atom on a superconducting Nb(110) substrate exhibits multi-orbital Shiba states which appear at different energies in spatially resolved STS maps [55,114] (Fig. 8(A)). Similar observations of the orbital structure of magnetic bound states have also been made for other types of transition metal species on different elemental superconducting substrates such as Pb [82,83], whereas Fe adatoms on superconducting Re(0001) only exhibit a single-orbital Shiba state [52]. For the case of individual Mn adatoms on superconducting Nb(110), four pairs of Shiba states appear within a hard gap of the superconducting substrate, as shown by the tunneling spectrum presented in Fig. 8(B). The coherence peaks of the superconducting Nb substrate (gray spectrum) appear at about $-1.5$ meV and $+1.5$ meV, while the spectrum within the gap appears featureless without the presence of a Mn adatom. If the tunneling spectrum is measured with the STM tip positioned above a Mn adatom, electron- and hole-like pairs of Shiba states with energies being symmetric around the Fermi level (corresponding to zero bias) show up. If another Mn adatom is now brought into vicinity of the first Mn adatom, hybridization of the Shiba states occurs [114,117]. The hybridization critically depends on the orientation of the Mn pair with respect to the crystallographic direction of the underlying Nb(110) substrate lattice (Fig. 8(A) right-hand side).

In Fig. 8(C) the evolution of the discrete Shiba states for an individual Mn adatom to Shiba bands of a close-packed linear Mn chain of variable length (up to 36 Mn atoms) built atom-by atom along the [001]-direction of the Nb(110) substrate is shown. The Shiba band formation occurs inside the gap of the superconducting Nb. Interestingly, some Shiba states cross the Fermi level for a chain length of about 5 Mn atoms. For larger Mn chain lengths, a small gap within the

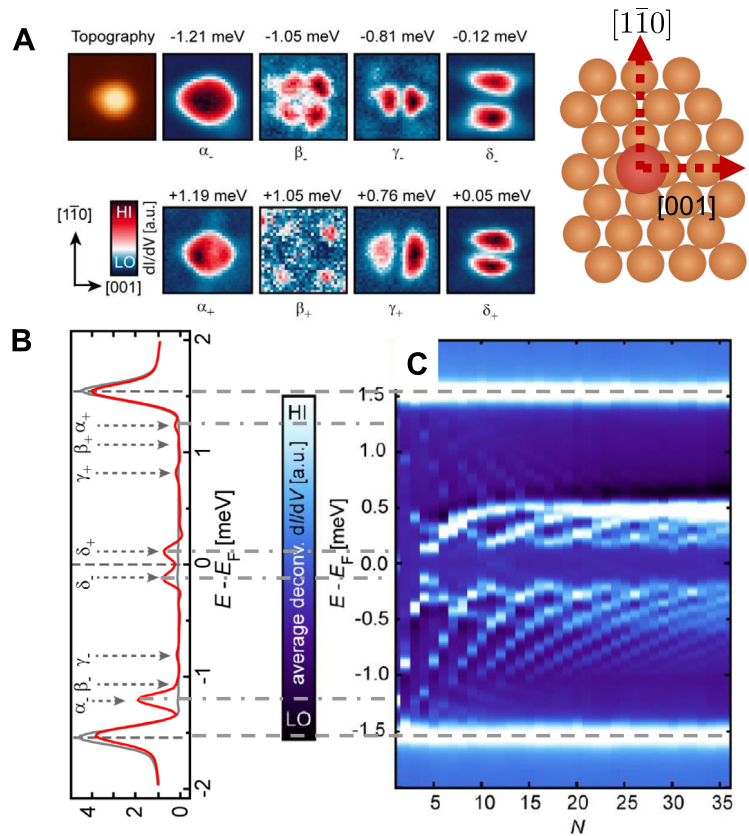

**Fig. 8.** From discrete Shiba states to a 1D Shiba band. (A) STM topography (top left) and differential tunneling conductance maps at different energies showing the spatial distribution of individual Shiba states of a multi-orbital Mn atom on a superconducting Nb(110) substrate. The crystallographic directions are indicated in the atomic structure model (right). [Taken from Ref. [114]]. (B) Local differential tunneling conductance spectrum measured above a single Mn atom adsorbed on the superconducting Nb(110) substrate (red). Four pairs of discrete Shiba states of this multi-orbital magnetic impurity can clearly be identified within the 'hard' superconducting gap of the Nb(110) substrate (gray). (C) Experimentally determined spectral evolution from discrete Shiba states of the individual Mn adatom to Shiba bands of Mn chains of increasing length (up to 36 atoms in [001]-direction) on a superconducting Nb(110) substrate. Individual Shiba states cross the Fermi level while a gap around zero energy is opened up within the Shiba bands forming with increasing chain length.
*Source:* Taken from Ref. [55].

Shiba bands on the order of 0.2 meV emerges. Whether this gap is associated with a topologically trivial or non-trivial phase of the Mn chain on Nb(110) hybrid system will be addressed in the next chapter.

It was found that linear close-packed Mn chains on superconducting Nb(110) behave very differently depending on the crystallographic direction along which they have been assembled [55,57,58,118]. Fig. 9 presents data of a close-packed linear chain consisting of 32 Mn atoms that have been assembled along the [1$\bar{1}$0]-direction with respect to the Nb(110) substrate lattice (Fig. 9(A)) using STM-based single-atom manipulation techniques. Fig. 9(B) (top part) shows the atomic-scale topography of that Mn chain, whereas spatially resolved STS maps at different energies are displayed below. At zero energy, enhanced spectral features at both chain ends clearly show up whereas the states at finite energies are spatially localized in the interior of the chain. Additional insight can be gained from the tunneling spectra measured at different locations as displayed in Fig. 9(C). The spectrum of the bare superconducting Nb(110) substrate (gray) exhibits again clear coherence peaks at −1.5 meV and +1.5 meV and a hard gap with no spectroscopic features in between. If the STM-tip is placed above the center of the Mn chain the tunneling spectrum (blue) clearly shows the emergence of in-gap Shiba states at finite energies. However, there are no states visible between −0.5 meV and +0.5 meV. However, if the STM-tip is placed above the chain's end a very clear zero-bias peak appears being energetically well separated from the finite-energy Shiba states. This observation is consistent with the presence of Majorana zero modes in such Mn chains on superconducting Nb(110), but it cannot be considered as a definite proof. Further experimental tests are required in order to probe additional characteristics of the zero-energy states which can also be found for other magnet–superconductor hybrid systems.

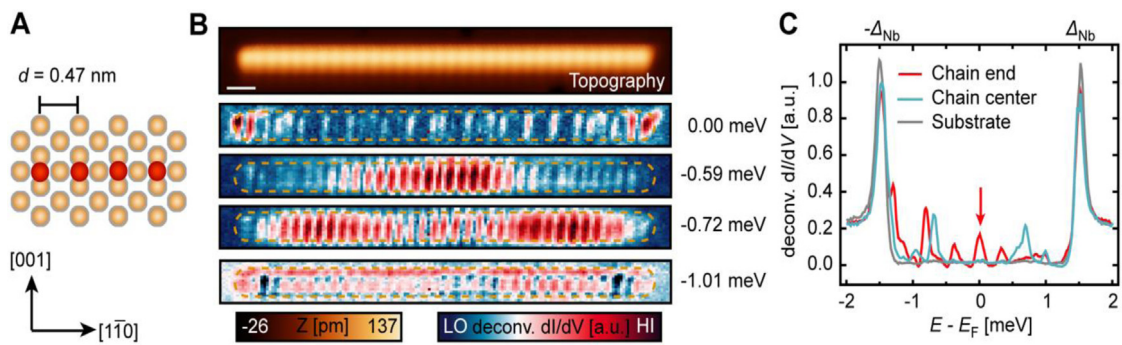

**Fig. 9.** Zero-energy modes in atomic Mn chains on superconducting Nb(110) being well separated from finite-energy Shiba states. (A) Schematic of a linear close-packed atomic Mn chain built atom-by-atom in [1$\bar{1}$0]-direction on a Nb(110) substrate. (B) STM topography of a 32-atom Mn chain built in [1$\bar{1}$0]-direction on superconducting Nb(110) and corresponding differential tunneling conductance maps at different energies. Zero-energy modes at both ends of the Mn chain are clearly visible. (C) Local dI/dV spectra measured at 350 mK, reflecting the LDOS of the superconducting Nb(110) substrate (gray), the LDOS at the center of the Mn chain (blue) as well as the LDOS at the chain ends (red). A zero-bias conductance peak is clearly visible at both ends of the Mn chain and energetically well separated from the finite-energy Shiba states within the 'hard' gap of the superconducting Nb(110) substrate.
*Source:* Taken from Ref. [57].

## 6. Probing the characteristics of Majorana quasiparticles in atomic-scale magnet–superconductor hybrid systems

In the following, we will present critical tests for the detection of Majorana quasiparticles in atomic-scale magnet–superconductor hybrid systems. This includes the observation of entangled zero-energy quantum states at both ends of the chains, the confirmation of the oscillatory behavior of the energies of hybridizing Majorana modes as a function of chain length, as well as the direct probing of the bulk-boundary correspondence of the topological superconducting state of atomic-scale magnet–superconductor hybrid systems.

A first critical test for the nature of end states is the observation of their behavior when an additional single atom is approached or added to only one end of the chain by using STM-based atom manipulation techniques. Since Majorana states should always show up at both ends of a chain being in a topological superconducting state, any asymmetric behavior of the spectroscopic features at both chain ends upon the modification of one chain end would indicate a trivial origin of the end states. Detailed studies on artificially constructed atomic Fe chains on Re(0001) have shown that local structural or chemical defects can significantly modify the electronic states at one end of the chains. For instance, a zero-bias feature at one chain end can be generated by a single-atom defect which, however, is then absent at the other end of the chain. Experiments involving single-atom control can therefore be extremely useful to probe the nature of zero-energy states mimicking Majorana bound states at the chain ends [52]. On the other hand, true Majorana bound states should always show up symmetrically at both ends of perfect, defect-free chains, as observed for pure close-packed linear chains of hcp Fe adatoms on a superconducting Re(0001) substrate [52].

Another insightful example is given by the behavior of short, close-packed Mn chains in [1$\bar{1}$0]-direction on a superconducting Nb(110) substrate [57] as presented in Fig. 10. In this case, individual Mn atoms are added one-by-one to an existing linear chain of 14 Mn atoms on Nb(110) (Fig. 10(A)). For the 14-atom long Mn chain, a zero-energy state is observed at both chain ends (Fig. 10(A), left). Interestingly, when a single Mn atom is added to one end of that chain, the zero-energy state becomes instantly split at both ends and appears at finite energy (Fig. 10(A), middle). If another Mn atom is added to one of the chain's ends, there is again a zero-energy state visible at both chain ends (Fig. 10(A), right). Since the spectroscopic signature at both chain ends are identical, though only one chain end is modified by the addition of a single Mn atom, it is clear that the low-energy spectroscopic feature arises from a quantum-coherent state over the full chain length, rather than from a localized, defect-related state [57].

By investigating the length-dependent behavior of the low-energy end states as a function of chain length, one finds a characteristic oscillating behavior (Fig. 10(B)) which can be nicely reproduced by theory (Fig. 10(C)) assuming that Majorana states from both ends of the relatively short chains hybridize. In fact, the energy splitting of the low-energy mode, as presented in Fig. 10(A) (middle), can be understood qualitatively and quantitatively by such hybridized Majorana modes which have also been named "precursor Majorana modes" [57]. It becomes clear from the length-dependent behavior of the low-energy states of Mn chains on Nb(110) (Fig. 10(B) and 10(C)) that longer chain lengths are required in this case in order to reduce the level of hybridization of Majorana modes from both chain ends and to get into the regime of topologically protected Majorana bound states. Interestingly, this regime can be reached for much smaller chain lengths for the case of linear close-packed hcp Fe chains on a superconducting Re(0001) substrate [52], as discussed together with Fig. 6(D). One possible reason could be the existence of a spin-spiral ground state for the close-packed hcp Fe chains on Re(0001) [52], presumably caused by the enhanced spin–orbit coupling, in contrast to the ferromagnetic ground state present for the Mn chains in [1$\bar{1}$0]-direction on Nb(110) [45].

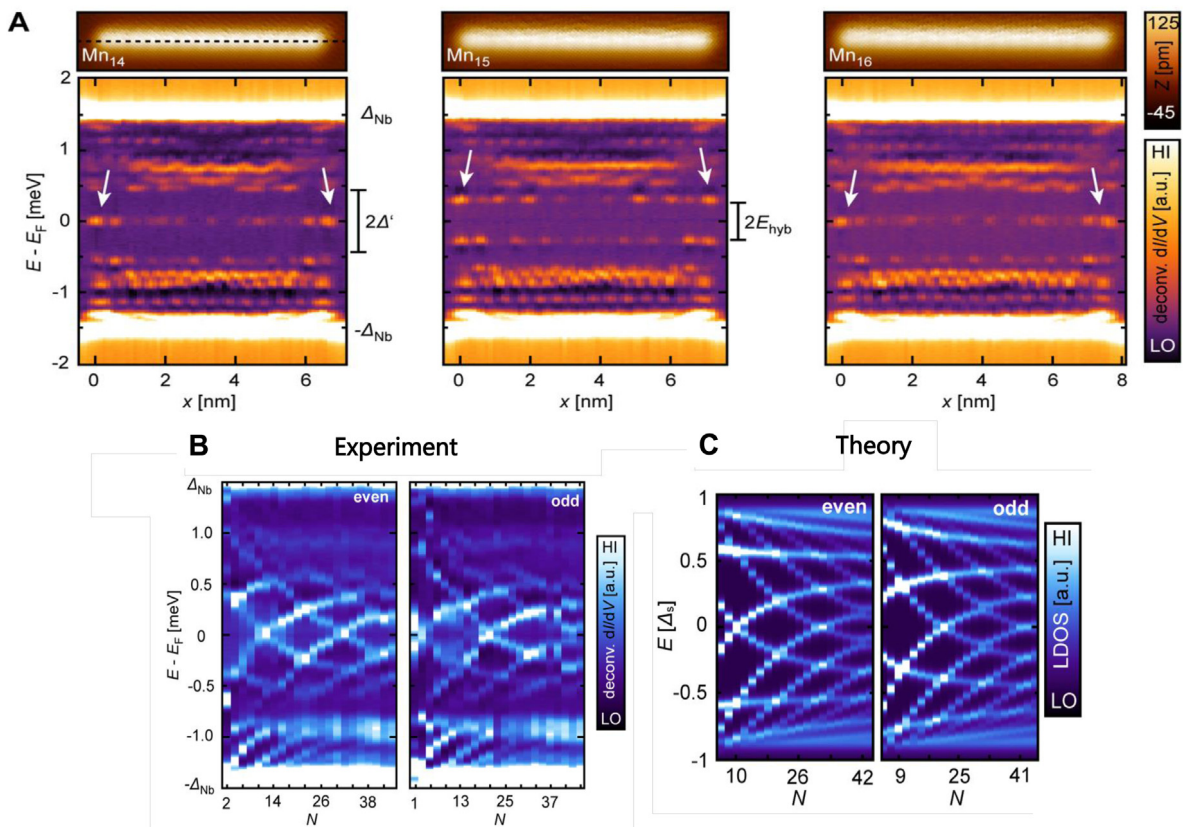

**Fig. 10.** Energy oscillations of hybridizing Majorana states as a function of Mn chain length. (A) For a 14-atom Mn chain built in [1$\bar{1}$0]-direction on a superconducting Nb(110) substrate, zero-energy modes at both chain ends being well separated from finite-energy Shiba states can be observed by spatially and energy-resolved STS measurements (left). By adding a single Mn atom at one end of the Mn chain, the end state splits in energy in a correlated fashion at both chain ends as characteristic for hybridizing Majorana modes (middle). By adding another Mn atom, the Majorana modes appear again at zero energy at both ends of the chain (right). (B) Energy of the sub-gap states as a function of Mn chain length for an even (left) or odd (right) number of atoms in the chain, as determined experimentally by STS. An oscillatory type of behavior as a function of chain length is clearly observed. (C) This oscillatory behavior can be reproduced by theoretical model-type calculations assuming strong hybridization in relatively short magnetic chains proximitized to a superconducting substrate.
*Source:* Taken from Ref. [57].

In addition to probing the quantum-coherent nature of Majorana states and the theoretically predicted oscillating behavior of the energies of hybridizing Majorana modes as a function of chain length [57], there is a much more direct way to prove the existence of Majorana modes. Since Majorana zero-energy modes are a direct consequence of a topological superconducting state, the ultimate experimental approach is the measurement of a topologically non-trivial band structure of the magnet–superconductor hybrid system. This has to be done on a single chain with atomic-scale spatial resolution and micro-eV energy resolution at sub-Kelvin measurement temperature. In contrast to topological insulators, for which topological band structures are routinely determined by angle-resolved photoemission spectroscopy (ARPES), this technique is not applicable for measuring the sub-gap band structure of an atomic-scale magnet–superconductor hybrid system. However, this ultimate goal can be achieved by Fourier-transform scanning tunneling spectroscopy (FT-STS) as explained in the following.

Fig. 11 presents the general principle of probing the sub-gap band structure of a topological superconductor by Bogoliubov quasiparticle interference mapping based on STS experiments. In Fig. 11(A) a topologically non-trivial band structure is schematically shown. By changing the energy, different k-states can be probed in reciprocal space. However, STS is a real-space technique. Nevertheless, information about the sub-gap band structure of the investigated magnet–superconductor hybrid systems can still be obtained: In an atomic chain of finite length, the electronic states (in case of a superconducting system, these are in fact Bogoliubov quasiparticle states) are back-reflected at the chain's ends so that $k$- and $-k$-states interfere with each other, thereby forming standing waves (Fig. 11(B)). These standing waves in real space exhibit an energy-dependent wavelength. The latter can directly be probed by STS, while the corresponding band dispersion can be obtained by subsequent Fourier transformation of the STS data sets. The possibility of measuring the band structure of a topological superconductor allows for a direct probing of the 'bulk-boundary correspondence' of topologically non-trivial zero-energy Majorana modes.

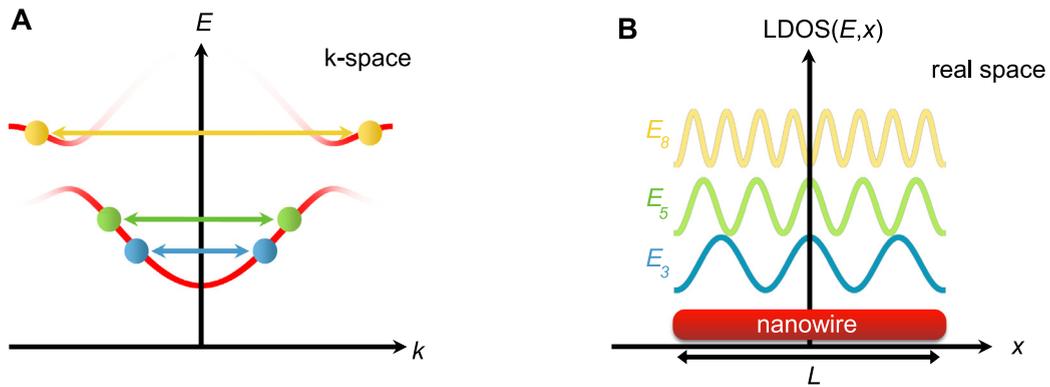

**Fig. 11.** Probing the sub-gap band structure of a topological superconductor. (A) Topologically non-trivial band structure in *k*-space. By changing the energy of the states, different *k*-states can be probed in reciprocal space. (B) In an atomic chain of finite length, the electronic states are back-reflected at the chain's ends so that $k$- and $-k$-states interfere with each other, thereby forming standing waves. These standing waves in real space exhibit an energy-dependent wavelength. The latter can directly be probed by STS, while the corresponding band dispersion can be obtained by subsequent Fourier transformation of the STS data. Thereby, the 'bulk-boundary correspondence' of topologically non-trivial zero-energy modes (Majorana states) can be probed experimentally.

An example for a topological band structure determination by STS mapping at sub-Kelvin temperatures is given by close-packed linear Mn chains in [001]-direction on a superconducting Nb(110) substrate. Such Mn-chains exhibit a FM ground state [45]. The emergence of a gapped Shiba band with increasing Mn chain length has already been discussed together with Fig. 8. In order to determine whether the observed gap of about 0.2 meV is a result of a topologically non-trivial superconducting state, STS mapping of Bogoliubov quasiparticle interference in Mn chains of variable length, up to about 50 Mn atoms, has been performed [55,58]. Fig. 12(A)–(C) presents three STS data sets of close-packed [001]-chains consisting of 17, 20 and 26 Mn atoms. The sub-gap band structure, i.e. the states inside the gap between $-1.5$ meV and $+1.5$ meV of the superconducting Nb(110) substrate, is clearly resolved and exhibits a mini-gap of about 0.2 meV, consistent with the data set presented in Fig. 8. The energy-dependent interference patterns on both sides of the mini-gap, corresponding to electron- and hole-like states, are nicely visible. Their evolution as a function of chain length can be studied as atom-by-atom is attached to the pre-existing Mn chain by STM-based single-atom manipulation [55]. Note that the atom at the foremost STM-tip apex should never change, otherwise the STS data sets obtained for different chain lengths could not be compared with each other. Furthermore, all Mn chains of different length have to be defect-free, otherwise the method of Bogoliubov quasiparticle interference would not lead to the desired goal, namely the determination of the sub-gap band structure of a disorder-free magnet–superconductor hybrid system.

Based on the exceptional quality of the STS data sets presented in Fig. 12(A)–(C) as well as for other Mn chain lengths, a Fast Fourier Transform (FFT) yields information about the sub-gap band dispersion for Mn chains of variable length [55]. An example is presented in Fig. 12(D) where the resulting band structure of a disorder-free 34-atom long Mn chain on superconducting Nb(110) is displayed. The observed band dispersion is fully consistent with the theoretical expectation for a topologically non-trivial superconducting state with a topological band gap of 0.18 meV (see Fig. 12(E)). Importantly, a non-zero spin–orbit coupling parameter has to be assumed in the calculations for the opening of the topological gap as expected [55]. In the presence of Rashba spin–orbit coupling, we can also rule out spin-degeneracy of the observed bands. Thus we can conclude that the gap is of topological origin. Moreover, there is a clear reminiscence to the spectrum of a gapped Dirac Hamiltonian. Based on a single-orbital Shiba-band model, one would expect to observe zero-energy Majorana bound states at the chain ends, as shown in Fig. 12(F). However, we already know that the assumptions of the most simple 1D theoretical models are not fulfilled for the case of Mn chains on superconducting Nb(110): first, the Mn adatoms exhibit multi-orbital Shiba states (in contrast to the Fe adatoms on superconducting Re(0001)), as discussed together with Fig. 8. Second, real multi-orbital systems will not exhibit only a single low-energy band. Third, real magnet–superconductor hybrid systems are not perfect 1D systems due to the 3D nature of the superconducting substrate, as can be seen by comparison of truly 1D models and models for atomic spin chains on 2D substrates [119].

Indeed, unexpected spatial distributions of the zero-energy states of linear, close-packed FM Mn chains in [001]-direction on superconducting Nb(110) were found experimentally [55,58]. Fig. 13(A) shows the STM topography as well as energy-resolved STS maps for the 34-atom long Mn chain, for which the topologically non-trivial sub-gap band structure has been measured before (Fig. 12(D)). The zero-energy mode does not appear as end states, but exhibits side features with respect to the spatial location of the Mn chain (Fig. 13(A)). The shift of the spectral weight off the chain is the reason why there is no visible spectral weight in the spectra shown in Fig. 12(A)-(C) which are measured along the center positions of the adatoms. In contrast, all finite-energy Shiba states are localized in the interior of the Mn chain. Theoretical calculations of the topological phase diagram (Fig. 13(B)) confirms that the hybrid Mn [001]-chain on superconducting Nb(110) system is in a topologically non-trivial superconducting state, in qualitative agreement with the measurement of a topologically non-trivial sub-gap band structure. Moreover, the theoretically derived spatial distributions of the zero-energy Majorana modes within the topologically non-trivial superconducting phase for different Mn chain lengths are in quantitative agreement with the experimentally determined zero-energy STS maps for different Mn chain lengths [58] (Fig. 13(C)). Localized MZMs could still be observed by reducing the coherence length or by building significantly longer chains. Similar "side features" of Majorana zero-energy modes have also been found experimentally and theoretically for other multi-orbital magnet–superconductor hybrid systems, such as atomic Fe chains on superconducting Nb(110) [58]. Model-type calculations for more realistic magnet–superconductor hybrid systems are therefore of utmost importance in order to explore Majorana physics for real materials and for guiding future experimental efforts based on the atomic magnetic chain on superconductor platform.

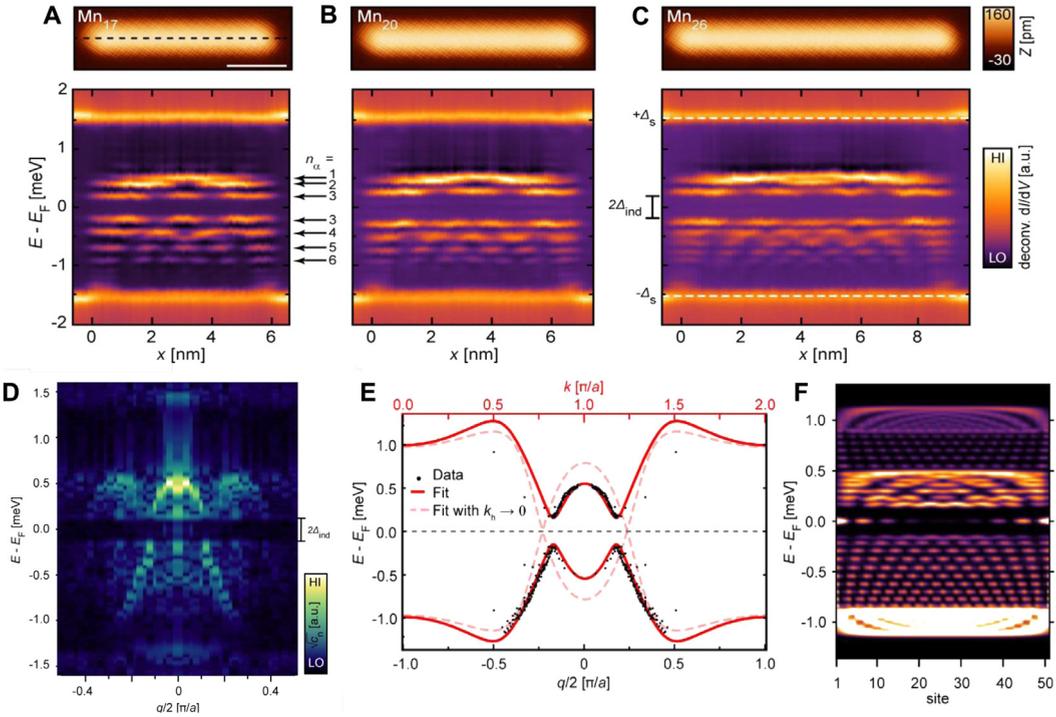

**Fig. 12.** Emergence of a topological Shiba band in atomic Mn chains of variable length built atom-by-atom along the [001]-direction on a superconducting Nb(110) substrate. (A-C) STM topography of a 17-atom (A, top), 20-atom (B, top), and 26-atom (C, top) Mn chain in [001]-direction on superconducting Nb(110) as well as the corresponding spatially and energy-resolved differential tunneling conductance maps (below), measured along a path at the top of the chains (dotted line in A). The spatially periodic Shiba states at finite energies and a well pronounced gap around zero energy are clearly visible. (D) By Fourier transformation of the real-space STS data, the dispersion of the sub-gap Shiba bands can be derived. (E) The experimentally determined dispersion of the dominant Shiba band (black data points) can very well be fitted by a topologically non-trivial Shiba band of a magnet–superconductor hybrid system with finite spin–orbit coupling parameter $k_h$ (red curve). In the limit of vanishing spin–orbit coupling, the topological gap would close (light red curve). (F) Calculated spatially and energy-resolved LDOS map of a topologically non-trivial Shiba band. Zero-energy Majorana modes at both chain ends are expected to be observable, which is, however, in apparent disagreement with the experimental results (A-C).
*Source:* Taken from Ref. [55].

In conclusion, early theoretical toy models for predicting topological superconductivity and associated Majorana zero modes in quasi-1D magnet–superconductor hybrid systems fail to explain the strong dependence of the emergence of topological phases on the microscopic details, such as the crystallographic direction of the magnetic atom chains with respect to the underlying atomic lattice of the superconducting substrate, the particular type of magnetic atoms, as well as the distance between them. While ferromagnetic close-packed Mn chains orientated along the [001]- and [1̄10]-directions on superconducting Nb(110) substrates have clearly revealed the signatures of topological superconductivity and associated Majorana zero modes, antiferromagnetic close-packed Mn chains in [1̄11]-direction on Nb(110) exhibit topologically trivial end states, which has been proven by introducing a locally perturbing defect on one of the chain's ends [118]. In general, the methodology of perturbing edge modes locally is a powerful tool to probe their stability against local disorder.

Dilute magnetic chains made up of Cr atoms with some larger interatomic spacing have been built on superconducting Nb(110) substrates as well [59]. Independent of the crystallographic orientation of these dilute Cr atom chains (in [001]-, [1̄11]- or [1̄10]-directions) on Nb(110), only topologically trivial end states could be identified experimentally, in agreement with multi-orbital tight-binding calculations [59]. Like the antiferromagnetic close-packed Mn chains on Nb(110), these dilute Cr chains on the same superconducting substrate provide further examples that zero-energy end states of trivial origin can mimic signatures of Majorana zero modes. Therefore, additional experimental methods for their unambiguous detection are required, such as probing the bulk–boundary correspondence or experiments with local perturbations of the end states.

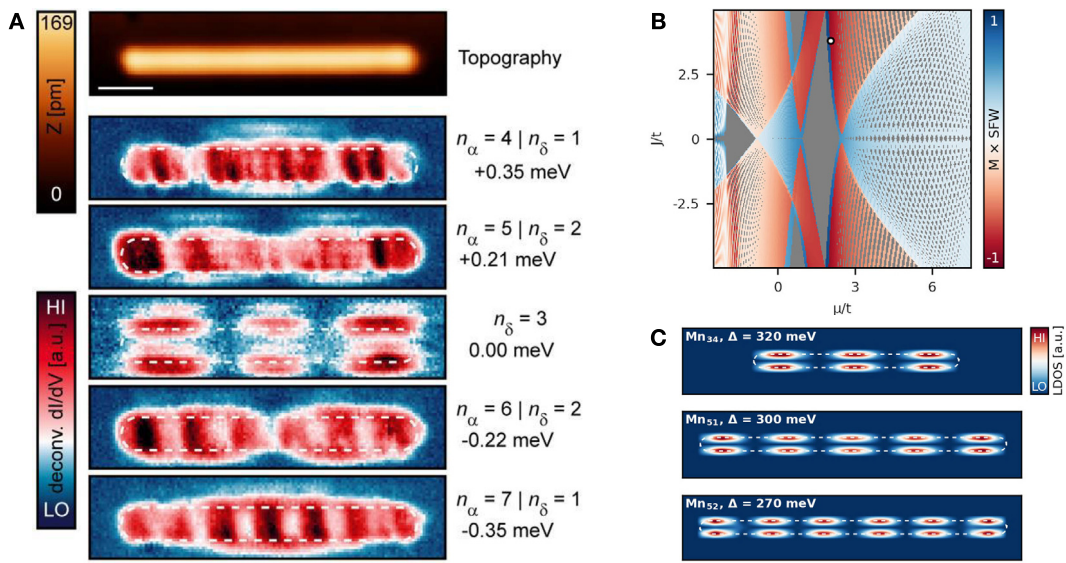

**Fig. 13.** Unusual spatial distribution of zero-energy Majorana states. (A) STM topography (top) and corresponding differential tunneling conductance maps at various energies (below) for a 34-atom Mn chain built atom-by-atom along the [001]-direction on a superconducting Nb(110) substrate. The zero-energy modes do not show up as end states, but rather as 'side features' with respect to the location of the Mn chain (white dotted lines). [Taken from Ref. [55]]. (B) Theoretical calculation of the topological phase diagram for a Mn chain on superconducting Nb(110) revealing large regions of the topologically non-trivial phase. (C) Theoretically derived spatial distributions of the zero-energy modes in the topologically non-trivial regime for Mn chains of variable length oriented along the [001]-direction with respect to the Nb(110) lattice. The theoretically calculated side features of the $Mn_{34}$-chain, which can be assigned to Majorana modes based on the calculated topological invariant, closely resemble the experimentally observed spatial pattern of the zero-energy state in (A).
*Source:* Taken from Ref. [58].

Let us summarize our discussion of whether or not Majorana quasiparticles were observed. In Fe-chains on Re(0001) the observed zero-energy states are compatible with MZMs. Also ab-initio based theory simulations support this interpretation. However, excited bulk states overlap with these zero-energy states due to the limited energy resolution of the STS measurements at 300 mK. To observe isolated MZMs inside a hard gap would require a reduction of temperature by a factor of 10.

Mn[1$\bar{1}$0]-chains on Nb(110) also feature zero-energy states for certain chain lengths. The experimental measurement of the low-energy states as a function of chain length revealed, however, an oscillatory behavior which can be understood through hybridized MZMs. Longer chains would be required to obtain stable, isolated MZMs. In contrast, Mn[001]-chains on Nb(110) were shown to possess a gap within the Shiba bands which is reminiscent of a gapped Dirac Hamiltonian, suggesting the gap is of topological origin. The associated zero-energy states do not possess significant spectral weight at the chain ends. Instead, they feature a standing-wave pattern with spectral weight pushed to the sides of the chain. Extensive (ab-initio based) theoretical modeling reproduced such zero-energy states, and identified them as strongly hybridized MZMs. Chain lengths of several hundred atoms would be required to obtain stable, isolated MZMs. Probably the same applies to Fe[001]-chains on Nb(110), although this system was only little studied.

While superconducting Nb substrates offer the largest superconducting gap size among all elemental superconductors, thereby facilitating the observation and separation of in-gap states arising from the interaction of the superconductor with magnetic adatoms and chains, other superconducting substrates, such as binary alloys, should seriously be considered as well. In particular, superconductors including Bi or Pb offering high spin–orbit coupling might be favorable in view of a possible optimization of the topological gap size and therefore the robustness of emerging Majorana zero modes. An interesting candidate is the $\beta$-phase of $Bi_2Pd$ which has also been used as substrate for atom-by-atom fabricated Cr chains in different crystallographic directions [56]. Fig. 14 shows STM images of two different types of Cr chains, a close-packed one built along the [110]-direction (Fig. 14(A)) and another one built in [100]-direction with a double spacing between the Cr adatoms (Fig. 14(B)). The evolution of the differential tunneling conductance curves as measured at the end atoms from dimers up to tetramers is displayed in Figs. 14(C) and (D). As visible in Fig. 14(C), as few as four Cr atoms are sufficient to close the gap due to the evolution of Cr-induced in-gap states. Theoretical calculations showed that this type of Cr chain in [110]-direction exhibits a ferromagnetic ground state and that a sizable Rashba coupling of the $\beta$-Bi$_2$Pd surface can lead to a topological phase transition for Cr chain lengths as short as eight atoms. Unfortunately, only 4-atom long Cr chains could be built on this particular superconducting substrate being too short for Majorana end states to emerge. On the other hand, the short Cr chains built in [100]-direction showed a persistent gap being rather independent of chain length (Fig. 14(D)). Theoretical calculations have proven that this type of Cr chain will not induce a topological phase transition on the superconducting $\beta$-Bi$_2$Pd surface, thereby providing another example of the strong dependence of the emergence of topological superconductivity on the microscopic details of magnet–superconductor hybrid systems.

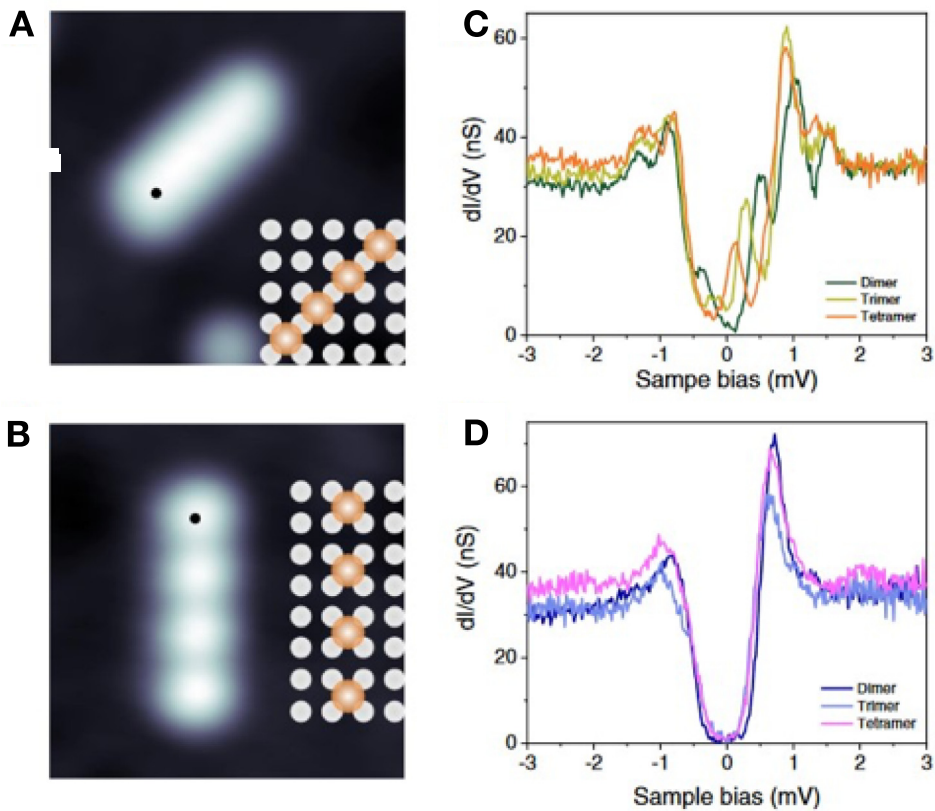

**Fig. 14.** Atom-by-atom fabricated Cr chains built on a $\beta$-Bi$_2$Pd superconducting substrate. (A) Topographic STM image and corresponding structure model of a close-packed 4-atom long Cr chain built in [110]-direction. (B) STM image and structure model of a 4-atom long Cr chain built along the [100]-direction with a double spacing between the Cr atoms. (C,D) Differential tunneling conductance spectra measured at the end atoms for dimer, trimer, and tetramer Cr chains built in the two different crystallographic directions.
*Source:* Taken from Ref. [56].

## 7. Towards braiding of Majorana states in artificially constructed atomic-scale networks

After discussing critical tests for the observation of Majorana zero modes in 1D magnet–superconductor hybrid systems, the successful atomic-level fabrication of various model-type patterns and network structures as basis for demonstrating the controlled manipulation of zero-energy Majorana bound states, including their creation, annihilation, adiabatic motion, and braiding will be presented in this chapter. Various concepts of Majorana state manipulation will be discussed and evaluated with respect to potential future applications.

Topological quantum computing based on fusion and braiding of non-Abelian anyons can be fault-tolerant, i.e., error-free. As explained in Box 1, the two MZMs at the ends of a Kitaev or Shiba chain constitute a physical quantum bit. By exchanging them once, a so-called *braiding phase* of $\pi/2$ is generated (as compared to a $\pi$ (0) phase for fermions (bosons)), constituting the case of fractional statistics [120,121]. Exchanging the two MZMs a second time brings them back to their

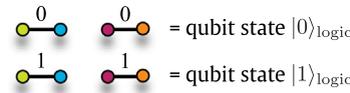

**Box 4: Braiding of Majorana zero modes**

(I) The two ground states of a Kitaev chain can be distinguished by their fermion-parity: $|0\rangle$ (even) and $|1\rangle$ (odd). In order to keep the total fermion parity constant, a second physical quantum bit is required to build one logical quantum bit.

Example 1:
1 quantum bit,
total parity even

= qubit state $|0\rangle_{\text{logic}}$
= qubit state $|1\rangle_{\text{logic}}$

Example 2:
2 quantum bits
(dense encoding),
total parity odd

= $|0,0\rangle_{\text{logic}}$
= $|1,1\rangle_{\text{logic}}$
= $|1,0\rangle_{\text{logic}}$
= $|0,1\rangle_{\text{logic}}$

The two examples nicely illustrate how the additional physical quantum bit is only used to conserve the total fermion parity.

(II) To understand the connection between braiding and quantum logic gates, consider the quantum bit from Example 1.

Even parity basis: $\begin{pmatrix} |0\rangle_{\text{logic}} \\ |1\rangle_{\text{logic}} \end{pmatrix} \hat{=} \begin{pmatrix} |0,0\rangle_{\text{physic}} \\ |1,1\rangle_{\text{physic}} \end{pmatrix}$ with definitions $|0,0\rangle_{\text{physic}} = |\text{vac}\rangle$ and $|1,1\rangle_{\text{physic}} = d_1^\dagger d_2^\dagger |\text{vac}\rangle$

The braid operator $\mathcal{B}_{23}$ which exchanges the "blue" ($\gamma_2$) and the "red" MZM ($\gamma_3$) in Example 1 is defined as $\mathcal{B}_{23} = \exp\left(\frac{\pi}{4}\gamma_3\gamma_2\right) = \frac{1}{\sqrt{2}}(1+\gamma_3\gamma_2) = \frac{1}{\sqrt{2}}\left(1 + id_2 d_1^\dagger + id_2^\dagger d_1^\dagger - id_2 d_1 - id_2^\dagger d_1\right)$.

In the even parity basis, $\mathcal{B}_{23} = \frac{1}{\sqrt{2}}\begin{pmatrix} 1 & -i \\ -i & 1 \end{pmatrix}$. Performing $\mathcal{B}_{23}$ twice, one finds $X = \begin{pmatrix} 0 & 1 \\ 1 & 0 \end{pmatrix} = i\mathcal{B}_{23}^2$.

(III) 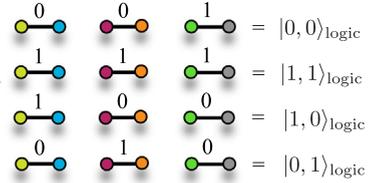

Z gate | X gate | Hadamard gate | Controlled Z gate
1-Qubit Gates | 2-Qubit Gate

original position. Specifically, if the system is initially prepared in the even parity ground state $|0\rangle$, the double-exchange of the two MZMs does not alter the state, $|0\rangle \to |0\rangle$. If, however, the initial state is the odd parity state $|1\rangle$, i.e., the Majorana bound state is occupied, then the double-exchange of the two MZMs results in a minus sign for the wavefunction, $|1\rangle \to -|1\rangle$. In matrix notation, the exchange corresponds thus to a Pauli Z matrix or a "Z gate". As demonstrated in Box 4, two pairs of MZMs (i.e., two chains) are required to realize a logical quantum bit. By exchanging ("braiding") $\gamma_2$ and $\gamma_3$ (see Box 4) the quantum bit which initially was in state $|0\rangle$ can be brought to an equal-superposition state $\sim(|0\rangle - i|1\rangle)$. By exchanging them again, one obtains the state $\sim |1\rangle$. The entire exchange process corresponds to the dynamical flipping of the quantum bit, realizing an X gate. As proposed theoretically [30,31], a T-junction or a Y-junction provides the simplest model-type platform enabling an adiabatic exchange and an unambiguous demonstration of the braiding of Majorana bound states, thereby testing the theoretically predicted non-Abelian quantum exchange statistics of Majorana quasiparticles. By preparing four MZMs on a T-junction (Fig. 15(A)), an X-gate can be performed via braiding: in Fig. 15(B) the transition amplitudes from the initial many-Majorana wave functions $|0\rangle$ and $|1\rangle$, respectively, to the time-evolved, braided states are shown. Similarly, other quantum logical gates can be performed, and in particular all Clifford gates can be realized via braiding [122].

Over the past decade, pioneering theory work about dynamic braiding simulations with MZMs was reported. Exact diagonalization remains an important method [62,124,125] which even allows the inclusion of arbitrary electron–electron interactions; its limitation clearly is the Hilbert space dimension, and only very short system sizes can be realized which hinders the investigation of four or more MZMs. To restrict the investigation to single-particle states is a common practice, see for instance Refs. [126,127]; similarly, the study of low-energy effective theories is another approach to braiding simulations [128–131]. Both approaches typically ignore bulk states and their influence on the Majorana ground-state manifold. A promising approach is to time-evolve the superconducting quasiparticles as done in Refs. [132–134]. It remains an interesting question whether such an approach can be used to simulate key observables in a many-body framework, i.e., with inclusion of additional (finite-energy) quasiparticles in addition to the MZMs. Simulations based on the Onishi formula [135] work well unless one encounters a sign problem [136,137]. The two methods which seem to perform without any issues are the *covariance method* [138–141] and the *time-dependent Pfaffian method* [60,142].

The mathematical framework of braiding anyons is well-established [122,143–148], and the connection to physical systems has been made. What is required to bridge these works and future experiments are full-scale many-body simulations to study the braiding dynamics of multiple Majorana-based quantum bits. First steps in this direction have been taken [60,63,124,140]. Going beyond a single qubit, comprised of two pairs of MZMs, leaves the choice to work with $M+1$ pairs of MZMs (dense qubit encoding) or with $2M$ pairs of MZMs (sparse qubit encoding) to constitute $M$ logical qubits [26,28]. The most promising strategy is to use the sparse qubit encoding for all single-qubit operations, and in order to entangle two qubits (i.e., to perform any controlled 2-qubit gate) one would switch to the dense encoding via projective measurements. Recently, a CNOT gate was demonstrated numerically on a system consisting of two quantum bits using the dense qubit encoding [123,149], as an example for multi-quantum bit entanglement [60]. The corresponding world lines for the six MZMs on a multi T-junction (Fig. 15(C)) lead to transition amplitudes shown in Fig. 15(D): if and only if the control quantum bit is active (i.e., states $|0,1\rangle$ and $|1,1\rangle$) the CNOT gate flips the target quantum bit (i.e., resulting in states $|1,1\rangle$ and $|0,1\rangle$). This numerical simulation illustrates the prospect of actual computing with MZMs.

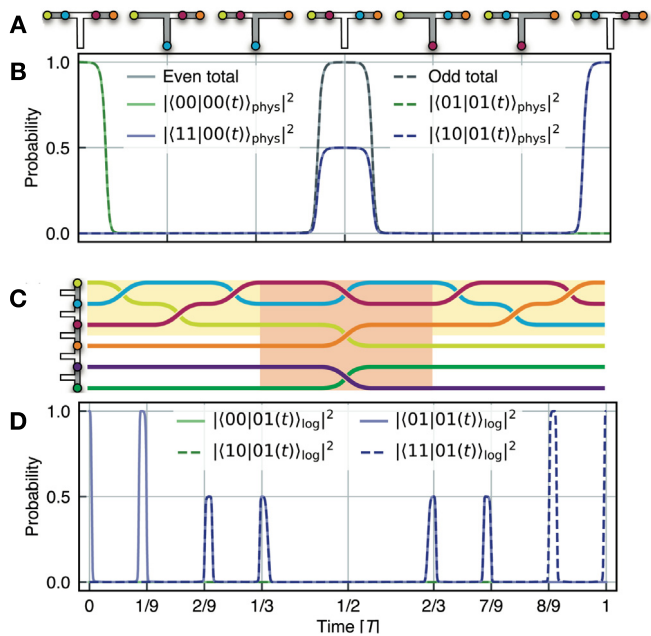

**Fig. 15.** Simulations of braiding multiple Majorana zero modes on tri-junctions. (A) Four Majorana modes are initially placed on a T-junction (gray areas are topological, white areas trivial). By changing the topological segments, the MZMs are moved on the T-junction, allowing the exchange of the two inner Majorana zero modes. (B) Corresponding transition amplitudes: two physical quantum bits with states $|0,0\rangle_{phys}$ and $|1,1\rangle_{phys}$ correspond to the logical quantum bit states $|0\rangle_{logic}$ and $|1\rangle_{logic}$, respectively. (C) Braiding protocol for six Majorana zero modes placed on a multi T-junction geometry, realizing a CNOT or CX gate [123] (yellow and orange regions correspond to Hadamard and CZ gates, respectively). (D) Corresponding transition amplitudes of the four two-quantum bit states: if the second quantum bit, the *control* bit, is "1", an X gate is performed on the first one, the *target* bit.
*Source:* Taken from Ref. [60].

The fabrication of well defined complex magnetic adatom arrangements suitable for braiding cannot be achieved by self-assembly processes, but requires STM-based single-atom manipulation techniques with atomic-scale precision. T-junctions and Y-junctions (Fig. 16(A)) of magnetic adatoms on superconducting substrates have been proposed theoretically in order to demonstrate the braiding of Majorana bound states [30,31], as explained in the previous paragraph. The atomically precise fabrication of Y-junctions has been demonstrated for magnetic Fe adatoms on a superconducting Re(0001) substrate using STM-based single-atom manipulation [150] (Fig. 16(B)). Moreover, artificially constructed atomic Fe chain networks on superconducting Re(0001) [150] (Fig. 16(C)) can serve as an interesting platform in order to test theoretical predictions of the behavior of Majorana bound states: At the points, where one, two, three, or four Fe chain ends meet, different signatures with respect to Majorana modes can be expected [61,151].

The challenge of moving MZMs in MSH structures has been first addressed in a proposal where magnetic adatom rings on conventional superconducting substrates can be prepared [62]. Such quantum corral structures can routinely be assembled atom-by-atom using a low-temperature STM [150,152,153]. It has been shown theoretically that a circular magnetic adatom arrangement on a conventional superconducting substrate can exhibit two topologically trivial and non-trivial segments if an in-plane magnetic field is applied (e.g. positioned in the center of the ring) [62]. Since the position of the four MZMs can be influenced by rotating the magnetic field, one should be able to move the Majorana bound states along the circle [62]. On the other hand, by changing the magnitude of the in-plane field, pairs of Majorana bound states can be created or fused.

Instead of globally manipulating the MSH structure and its MZMs in order to perform a quantum logic gate, an approach which allows the local addressability and tunability of the system is desirable. An interesting concept is based on using a combination of electron-spin resonance and scanning tunneling microscopy (ESR-STM) techniques [154–156]. Single atoms can be targeted and by applying controlled microwave pulses the spin-direction of an adatom can be changed. To rotate magnetic moments in opposite directions (the requirement for an antiferromagnetic alignment or, more generally, a non-coplanar spin structure) can be achieved by using different types of magnetic adatoms, or by changing the local magnetic structure [63,155,156]. A theoretical proposal has identified an adatom network on a superconducting substrate, where chain segments with out-of-plane ferromagnetic order realize a topological superconductor phase and where in-plane antiferromagnetic order lead to a gapped, but topologically trivial superconducting phase [63]. At the boundary points between topological and trivial segments MZMs are localized (Fig. 17(A)). By dynamically rotating the magnetic moments of adatoms, MZMs can be moved through the wire network enabling the realization of quantum logic gates such as Pauli Z and X gates [63]. A proposal for a $\sqrt{Z}$ gate, i.e., half a Z gate, is shown in Fig. 17(B)-(H) [63]. In particular, the world-lines of the braiding process were computed using a non-equilibrium density of states (Fig. 17(H)). These theoretical simulations demonstrate that the adiabatic manipulation of the magnetic structure of the adatoms can be sufficient to allow for the braiding of MZMs and computation of quantum logical gates in magnet–superconductor hybrid structures.

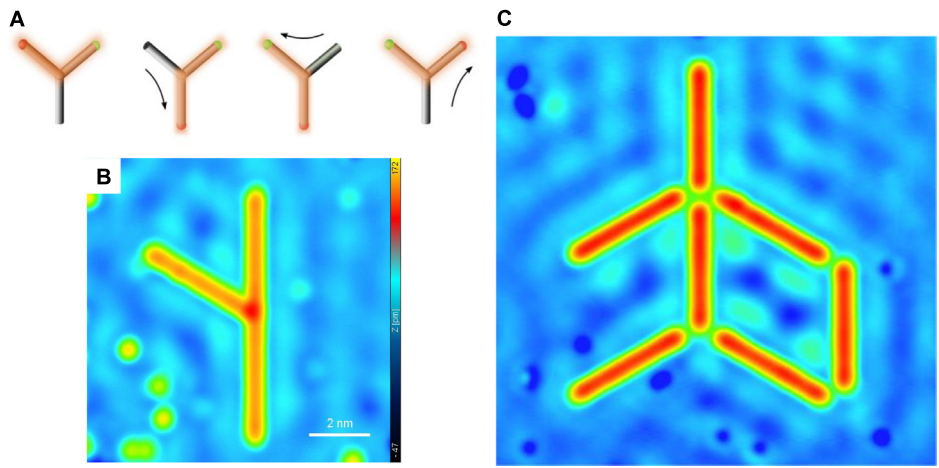

**Fig. 16.** Towards Majorana state manipulation in artificially constructed atomic-scale networks built atom-by-atom. (A) Theoretically proposed Majorana state manipulation ('braiding') in a tri-junction [31]. (B) Artificially fabricated tri-junction by sequential STM-based single Fe-atom manipulation on a clean superconducting Re(0001) substrate [150]. (C) Artificially constructed atomic Fe chain network on superconducting Re(0001) [150]. At the points, where one, two, three, or four Fe chain ends meet, different signatures with respect to Majorana modes can be expected [61].

As discussed previously for Mn-chains on Nb(110), spin chains in different crystallographic directions might feature different magnetic orders. Constructing Y and T junctions or even wire networks might thus result in systems where certain legs of a Y or T junction possess different magnetic orders. By changing these orders from out-of-plane to in-plane (as explained above), braiding might still be possible just like for magnetically homogeneous adatom chains [94].

A completely different route might be measurement-based braiding [157,158]. Originally proposed in the context of semiconducting wire–superconductor heterostructures [159], braiding of MZMs could be possible by a specific measurement process. Based on the insight that both the measurement and braiding relies on the parity operator, a technique has been developed which allows to braid MZMs *without* physically moving them. While there are no specific analogous proposals for MSH structures to date, such a strategy might also be of interest for the adatom chain–superconductor hybrid structures discussed in this review.

## 8. Conclusions and perspective

The amazing recent advances in the bottom-up fabrication of disorder-free atomic spin chains on superconducting substrates based on single-atom manipulation techniques have opened up new exciting possibilities for critical tests of Majorana physics. The current focus on Majorana state manipulation in perfect disorder-free model-type atomic-scale networks is motivated by the fact that alternative platforms for Majorana physics suffer from impurity-related effects which are difficult to control down to the atomic level. It has been theoretically estimated that the current level of disorder, e.g., in semiconductor Majorana nanowires is at least an order of magnitude higher than required for the emergence of topological Majorana zero modes [160]. In particular, charge impurities in semiconducting nanowires lead to serious complications regarding the realization and application of Majorana zero modes [161]. Disorder induced effects in metal-based Majorana platforms are expected to be even worse [160], therefore self-assembled magnetic chains on superconducting substrates with intrinsic disorder of various types are not a good alternative. Moreover, the fabrication of arbitrary model-type structures as basis for demonstrating Majorana state manipulation is almost impossible by using self-assembly techniques. On the other hand, atomically-precise single-atom manipulation on atomically clean superconducting substrates leads to disorder-free model-type platforms without any impurity- or disorder-related effects so that a detailed comparison between experimental and theoretical results becomes possible. Most importantly, based on recent advances in the study of Majorana zero modes in artificially fabricated magnet–superconductor hybrid structures, the next steps towards controlled Majorana state manipulation can be taken.

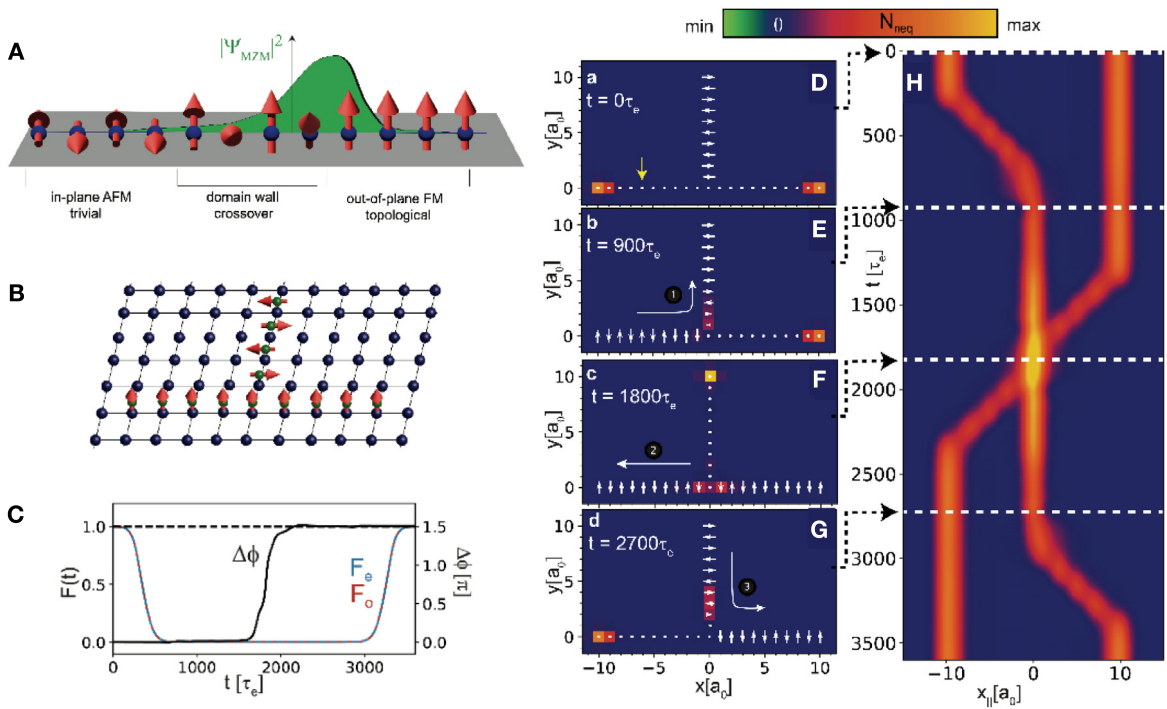

**Fig. 17.** Braiding of Majorana zero modes in magnet–superconductor hybrid structures by manipulation of the magnetic order of the adatoms. (A) Majorana mode at the boundary between a trivial superconductor (with in-plane AFM order) and a topological superconductor (with out-of-plane FM order). (B) Illustration of an upside-down T-junction with topological and trivial segments embedded in a 2D substrate. (C) Braiding dynamics of a $\sqrt{Z}$ gate: fidelity $F(t)$ indicating adiabaticity, i.e., absence of transitions into excited states, and the braiding phase $\Delta\phi$ reaching an odd-multiple of $\pi/2$. (D-G) Snapshots of the braiding dynamics, where the zero-energy non-equilibrium density of states is shown in color. (H) Non-equilibrium density of states projected onto a 1D axes vs time ($y$-axes) revealing the world-lines of the Majorana modes realizing the $\sqrt{Z}$ gate.
*Source:* Taken from Ref. [63].

An alternative bottom-up approach towards Majorana qubits based on chains of quantum dot–superconductor hybrids has recently been proposed [162]. In a first step, a two-site Kitaev chain using YSR states has been realized, showing signatures of so-called "poor man's Majorana states" with a gap larger than 0.07 meV. It has been argued that a two-site Kitaev chain realized by using YSR states should suffice for the demonstration of a prototypical Majorana qubit and the verification of non-Abelian properties based on fusion and braiding experiments.

Finally, it should be emphasized that the bottom-up approach for creating well-defined and disorder-free atomic spin chains on superconductors, as has been in the focus of this review, can straightforwardly be extended to two-dimensional magnet–superconductor hybrid systems, which are equally exciting in view of the realization of novel topological superconducting phases [163,164]. Indeed, two-dimensional Shiba lattices as a promising platform for topological superconductivity have recently been reported [165]. It has been highlighted that the possibility to design 2D lattices of different symmetries and lattice constants, combined with the use of distinct magnetic adatoms as elementary building blocks, offer great opportunities to artificially fabricate and control topologically protected superconducting states. Moreover, artificially built 2D lattices of magnetic adatoms on superconducting substrates allow for the investigation of quantum states as well as the interplay between symmetries and geometry of their boundaries.

In summary, the application of atomic-level fabrication and atomic-scale characterization techniques has opened up novel exciting directions towards the design of topological superconductivity in magnet–superconductor hybrid systems and associated Majorana zero modes. Theoretical concepts involving the adiabatic manipulation of the magnetism of the adatoms have been presented, which can potentially pave the way towards applications of Majorana quasiparticles for topological quantum computation.

## CRediT authorship contribution statement

**Stephan Rachel:** Writing – original draft, Visualization, Validation, Software, Resources, Methodology, Investigation, Funding acquisition, Formal analysis, Conceptualization. **Roland Wiesendanger:** Writing – original draft, Visualization, Validation, Supervision, Resources, Project administration, Methodology, Investigation, Funding acquisition, Formal analysis, Data curation, Conceptualization.

## Declaration of competing interest

The authors declare that they have no known competing financial interests or personal relationships that could have appeared to influence the work reported in this paper.

## Acknowledgments

We would like to thank our coworkers and collaborators Ph. Beck, L. Cornils, D. Crawford, T. Hodge, H. O. Jeschke, A. Kamlapure, H. Kim, E. Mascot, D. K. Morr, A. Palacio-Morales, K. Palotás, Th. Posske, L. Rózsa, L. Schneider, L. Szunyogh, M. Thorwart, and J. Wiebe for their contributions and for numerous discussions. Financial support from the European Union via the ERC Advanced Grant ADMIRE (project No. 786020), the Deutsche Forschungsgemeinschaft, Germany via the Hamburg Cluster of Excellence 'Advanced Imaging of Matter' (EXC 2056 - project ID 390715994) and the Australian Research Council, Australia through Grants No. DP200101118 and DP240100168 is gratefully acknowledged.